%% file: DYB_AD12_nim.tex
\documentclass[final,3p,times,twocolumn,longtitle]{elsarticle}

\journal{Nuclear Instrument and Method}

\usepackage{graphicx}
\usepackage{lineno}

\footnotetext[1]{Deceased}

\begin{document}

\begin{frontmatter}

\title{A side-by-side comparison of Daya Bay antineutrino detectors}

\input{dyb_authors_AD12.tex}

\begin{abstract}

\noindent The Daya Bay Reactor Neutrino Experiment is designed to determine precisely the neutrino mixing angle $\theta_{13}$ with a sensitivity better than 0.01 in the parameter sin$^22\theta_{13}$ at the 90\% confidence level.  To achieve this goal, the collaboration will build eight functionally identical antineutrino detectors.  The first two detectors have been constructed, installed and commissioned in Experimental Hall 1, with steady data-taking beginning September 23, 2011.  A comparison of the data collected over the subsequent three months indicates that the detectors are functionally identical, and that detector-related systematic uncertainties exceed requirements.

\end{abstract}


\end{frontmatter}

%\linenumbers

\section{Introduction}
\par
The precise determination of the neutrino mixing angle $\theta_{13}$ by the Daya Bay Reactor Neutrino Experiment (Daya Bay) requires measurement of the antineutrino flux from the six nuclear reactors at different baselines using eight antineutrino detectors~\cite{dyb}.  Detection of antineutrinos is via the inverse beta-decay (IBD) reaction:

\begin{equation}
\bar{\nu}_e + p \to e^+ + n\,. \label{eqn:IBD}
\end{equation}

\noindent The positron rapidly annihilates with an electron (prompt signal) while the neutron first thermalizes before being captured by a nucleus and releasing energy (delayed signal).

The value of $\sin^2{2\theta_{13}}$ can be determined by comparing the
observed antineutrino rate and energy spectrum with
predictions assuming oscillations.  The number of detected
antineutrinos $N_{\rm det}$ is given by
\begin{eqnarray}
N_{\rm det}&=&\frac{N_p}{4\pi L^2}\int{\epsilon\sigma P_{\rm sur}(E,L,\theta_{13})S dE}
\label{eq:absolute}
\end{eqnarray}
where $N_p$ is the number of free protons in the target, $L$ is the
distance of the detector from the reactor, $\epsilon$ is the
efficiency of detecting an antineutrino, $\sigma$ is the total cross
section of the IBD process, $P_{\rm sur}$ is the $\bar{\nu}_e \to \bar{\nu}_e$ survival
probability that depends on the value of $\sin^2{2\theta_{13}}$, and $S$ is the differential
energy distribution of the antineutrino.

With only one detector at a fixed baseline from a reactor, according
to Eq.~\ref{eq:absolute}, we must determine the
absolute antineutrino flux from the reactor, the absolute cross
section of the IBD reaction, and the efficiencies of
the detector and event-selection requirements in order to measure
$\sin^22\theta_{13}$.  It is a challenge to reduce the systematic uncertainties
of such an absolute measurement to sub-percent level, especially for
reactor-related uncertainties.

Mikaelyan and Sinev pointed out that the systematic uncertainties can
be greatly suppressed or totally eliminated when two detectors
positioned at two different baselines are utilized~\cite{Russian}.
The detector closer to the reactor core is primarily used to establish the
flux and energy spectrum of the antineutrinos. This relaxes the
requirement of knowing the details of the fission process and
operational conditions of the reactor.  In this approach, the value of
$\sin^22\theta_{13}$ can be measured by comparing the antineutrino
flux and energy distribution observed with the far detector to those
of the near detector.

According to Eq.~\ref{eq:absolute} for a single reactor core and single near and far detectors, the ratio of the number of antineutrino events with energy between $E$ and $E+dE$ detected at distance $L_{\rm f}$ (far detector) from the reactor core to that at a distance $L_{\rm n}$ (near detector) is given by
\begin{eqnarray}
\frac{N_{\rm f}}{N_{\rm n}}&=&
\left (\frac{N_{\rm p,f}}{N_{\rm p,n}}\right )
\left (\frac{L_{\rm n}}{L_{\rm f}}\right)^2
\left (\frac{\epsilon_{\rm f}}{\epsilon_{\rm n}}\right )
\left [\frac{P_{\rm sur}(E,L_{\rm f},\theta_{13})}{P_{\rm sur}(E,L_{\rm n},\theta_{13})}\right]
\end{eqnarray}

\noindent where $N_{\rm p,f}$ and $ N_{\rm p,n}$ refer to the number of target protons at the far and near sites, respectively.
The relative detector efficiency ($\epsilon_{\rm f} / \epsilon_{\rm n}$) can be determined more precisely than the absolute
efficiency.  Hence, the detector-related systematic uncertainty in this approach is greatly reduced.  Furthermore, the use of multiple modules at each site enables internal consistency checks.
Daya Bay will implement this strategy by deploying two functionally identical modules at each of two sites near the reactor cores, and four detectors at a site further away.

In this paper we will compare the performance of the first two antineutrino detectors (ADs) using three months of data.  The detectors were installed and commissioned side-by-side in the Daya Bay Experimental Hall (also known as EH1) during the summer of 2011.  The data used in this analysis were collected between September 23 and December 23, 2011.
To reduce the potential for biases during data analysis, Daya Bay has adopted a blind analysis. The baselines, the thermal power histories of the cores, and the target masses of the antineutrino detectors will be blinded. Before unblinding, nominal values for these quantities will be used.  An overview of the experimental site and detectors will be provided first, followed by a detailed comparison of the detectors' performance.

\section{Experimental site}

\par
The Daya Bay nuclear power complex is located on the southern coast of China, 55 km to the northeast of Hong Kong and 45 km to the east of Shenzhen. As shown in Fig.~1, it  consists of three nuclear power plants (NPPs), Daya Bay NPP, Ling Ao NPP, and Ling Ao-II NPP. The complex faces the sea on the southeast, and is adjacent to mountains on the northwest. Each NPP consists of two reactor cores. All six cores are functionally identical pressurized water reactors (Framatone M310 and its derivative CPR1000) of 2.9 GW thermal power~\cite{Guangdong}. The last core started commercial operation on Aug.\ 7, 2011. The distance between the cores for each pair is about 90 m. The Daya Bay cores are separated from the Ling Ao cores by about 1100 m, while the Ling Ao-II cores are around 500 m from the Ling Ao cores.

\begin{figure}[htb]
\begin{center}
\includegraphics[width=7.0cm]{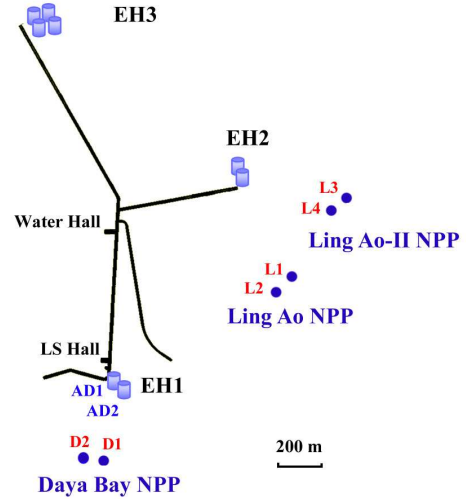}
\caption{Layout of the Daya Bay experiment. The dots are reactor cores, labeled as D1 and D2 for the Daya Bay NPP, L1 and L2 for the Ling Ao NPP, and L3 and L4 for the Ling Ao II NPP. Two antineutrino detectors, AD1 and AD2, are currently installed in the Daya Bay Experimental Hall (EH1).}
\end{center}
\end{figure}

\par
The Daya Bay experimental facility consists of surface facilities, three underground experimental halls, and two additional underground utility halls known as the Liquid Scintillator Hall (LS Hall) and the Water Hall. The surface facilities include a Surface Assembly Building (SAB) where the ADs are assembled, an office building, and a building housing the ventilation equipment for the underground halls.  Each detector hall contains a water pool instrumented to detect Cherenkov radiation, either two or four ADs installed inside the water pool, and modules containing four layers of resistive plate chambers (RPCs) over the top of the pool.  The LS Hall is the location used for producing and storing the liquid scintillator (LS) and gadolinium-doped liquid scintillator (Gd-LS), as well as for filling the ADs. The Water Hall is used to produce purified water for the water Cherenkov detectors. The underground halls are connected by horizontal tunnels with a 0.3\% slope to facilitate drainage of water.  The surface-access tunnel is 267 m long with a 10\% downward slope. The total length of the tunnel system is 3100 m. The tunnels are 6.2 m wide by 7.0 m high to allow for transportation of assembled ADs.

The mountain contour over the tunnels and experimental halls was surveyed prior to the ground-breaking of civil construction. An additional survey was completed later using GPS and Total Station technologies that determined the position of the detectors in each experiment hall with respect to the reactor cores to a precision of a few centimeters.  Based on these surveys, the Daya Bay Experimental Hall (EH1) has an overburden of 280 meter-water-equivalent (mwe) and is about 360~m from the center of the twin cores of the Daya Bay NPP.  The Ling Ao Experimental Hall (EH2) is 300 mwe deep and is about 500 m on average from the four cores of the Ling Ao NPP and Ling Ao-II NPP. The Far Hall (EH3) is 880 mwe deep, about 1910~m from the cores of the Daya Bay NPP and about 1540~m from the cores of the Ling Ao and the Ling Ao-II NPPs.  These approximate values are tabulated in Table 1.  The precise locations will be used in the final analysis.  At full thermal power, each AD in EH1 is expected to observe about 800 IBD events per day where the neutron is captured by a Gd nucleus.

\begin{table}[h]
\begin{center}
\begin{tabular}[c]{ccccc} \hline
  &Overburden & D.~B. & L.~A. & L.~A.~II\\\hline\hline
EH1 & 280 & 360 & 860 & 1310\\
EH2 & 300 & 1350 & 480 & 530\\
EH3 & 880 & 1910 & 1540 & 1550\\\hline
\end{tabular}
\caption{Approximate values for the overburden above and distances to the three experimental halls from the Daya Bay, Ling Ao, and Ling Ao-II NPPs. The overburden is in meter-water-equivalent and the distances are in meters.} \label{tab:baseline}
\end{center}
\end{table}

\section{Detectors}
\par
In each experimental hall there are three different kinds of detectors: the ADs, the water Cherenkov detectors, and the RPC detectors. In total, there will be eight 110-ton ADs, three water pools filled with 4400 t of purified water, and three arrays of RPCs covering a total of 800 m$^2$.

\subsection{Antineutrino detectors}

\par
Each AD has three nested cylindrical volumes separated by concentric acrylic vessels as shown in Fig.~2.  The outermost vessel is constructed of stainless steel and is known as the SSV.  The innermost volume holds 20 t of 0.1\% by weight Gd-LS that serves as the antineutrino target. The middle volume is called the gamma catcher and is filled with 21 t of un-doped liquid scintillator (LS) for detecting gamma-rays that escape the target volume.  The gamma-catcher increases the containment of gamma energy thus improving the energy resolution and reducing the uncertainties of the antineutrino detection efficiency. The outer volume contains 37 t of mineral oil (MO) to provide optical homogeneity and to shield the inner volumes from radiation originating, for example, from the photo-multiplier tubes (PMTs) or SSV.  Three automated calibration units (ACU-A, ACU-B, and ACU-C) are mounted at the top of the SSV.  Each ACU contains a LED as well as two sealed capsules with radioactive sources that can be lowered individually into the Gd-LS along either the centerline or inner edge, or in the LS.

\begin{figure}[htb]
\centering
\includegraphics[width=7cm]{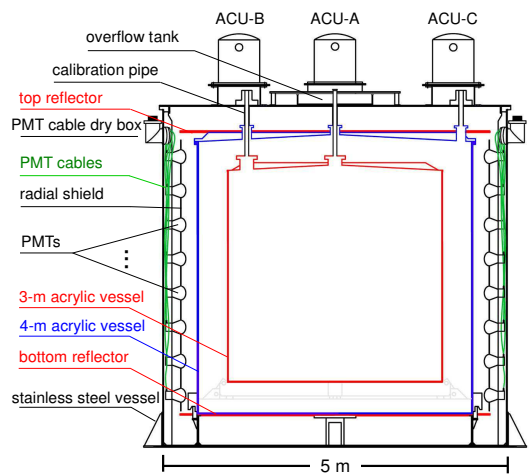}
\caption{Schematic for a Daya Bay antineutrino detector. }
\label{fig:AD_structure}
\end{figure}

\par
The acrylic vessel holding the Gd-LS has a diameter and height of 3.1 m and a wall thickness of 10 mm. This inner acrylic vessel (IAV) is nested within an outer acrylic vessel (OAV) with a diameter and height of 4 m and a wall thickness of 18 mm. Both the IAV and OAV are made of UV-transparent acrylic~\cite{bryce,band}. The lids are cone-shaped with a 3 degree tilt angle. There are two calibration pipes made of Teflon bellows connecting the IAV to the ACUs on the top of the AD. They are nested within larger Teflon bellows attached to the OAV. A third pipe located at the edge of the OAV provides access for ACU-C to the gamma catcher. Gd-LS and LS can flow along the calibration pipes to an overflow tank located at the top-center of the SSV.  The overflow tank is a nested two-layer acrylic vessel. The mineral oil has two separate overflow tanks also situated on the top of the SSV.  The SSV has a diameter and height of 5 m and a wall thickness of 12 mm. It is reinforced with ribs at the bottom, under the lid, and on the inner and outer surfaces of the barrel to provide sufficient mechanical strength for lifting after the AD is filled, to withstand the water pressure, and maintain an accuracy of 2 mm for the critical AD components and the location of calibration sources.  Two 4.5-m diameter, 2-cm thick reflective panels are placed at the top and bottom of the OAV to increase the photon-statistics and improve the uniformity of the energy response~\cite{ADproto}.  The reflectors are laminated with a film of ESR (Vikuiti$^{\rm TM}$ Enhanced Specular Reflector Film) sealed between two 1-cm thick acrylic panels. The ESR film has a reflectivity greater than 98\% across most of the relevant spectrum.

\par
There are 192 8-inch PMTs (Hamamatsu R5912) mounted on eight ladders installed along the circumference of the SSV and within the mineral oil volume. The PMT bulb is 20 cm from the OAV. To minimize non-uniformity of response, the PMTs are recessed in a 3-mm thick black acrylic cylindrical shield located at the equator of the PMT bulb.  A conical magnetic shield is wrapped around the dynode structure to protect the PMTs from stray magnetic fields.   The photocathodes operate at ground potential.  A single coaxial cable is used to supply positive high voltage and transmit the PMT signal to the front-end electronics.  The decoupling of the signal from the high voltage is performed inside the electronics room. The high-voltage system uses a CAEN SY1527LC mainframe.  Each mainframe houses eight A1932AP 48-channel high-voltage distribution modules. During data taking, the gain of the PMTs is set to $1\times10^7$.

\par
ACU-A sits on the central axis of the detector. ACU-B is located at a radius of 135.00 cm to calibrate and study edge effects within the IAV. ACU-C is located at a radius of 177.25 cm for calibrating the gamma catcher, on the opposite side to ACU-B. Each ACU is equipped with a LED, a $^{68}$Ge source, and a combined source of $^{241}$Am-$^{13}$C and $^{60}$Co. The Am-$^{13}$C source generates neutrons at a rate of 0.5 Hz. The rates of the $^{60}$Co and $^{68}$Ge sources are about 100 Hz and 15 Hz, respectively.  Since the AD is fully submerged in water, the ACUs are operated remotely.  The sources can be located to better than 0.5 cm along a vertical line down to the bottom of the acrylic vessels.  When not in use, the LED and sources are retracted into the ACUs, that also serve as shielding for the sources.

\par
Six 2-inch PMTs (Hamamatsu R7724) are installed at the top and bottom of the AD, to monitor the attenuation length of the Gd-LS and LS via optical windows on the reflective panels. A mineral oil clarity device is installed on the AD lid to monitor the attenuation length of the mineral oil by detecting blue LED light reflected back from a retroreflector at the bottom of the AD. The AD is also instrumented with two CMOS cameras (normally off) mounted at the top and bottom of a PMT ladder, and temperature sensors along the PMT ladders and in the overflow tanks. The liquid levels of the LS, Gd-LS, and MO within the overflow tanks are each monitored using redundant sensors (capacitance, ultrasonic, and cameras).

\par
All Gd-LS~\cite{gdls} and LS was produced in the LS Hall~\cite{Goett}.  The solvent of the LS and Gd-LS, linear alkylbenzene (LAB), was procured in one batch~\cite{LABproperty}. The fluor and wavelength-shifter are 3 g/L PPO and 15 mg/L bis-MSB, respectively.  Gadolinium chloride reacts with 3,5,5-trimethylhexanoic acid (TMHA) to form a solid complex Gd(TMHA)$_3$ that dissolves in LAB.  The Gd-LS was produced in 4-ton batches and stored in five 40-ton acrylic storage tanks, each with internal circulation systems.  Four tons of Gd-LS are taken from each of the five storage tanks and pre-mixed prior to the filling of each AD to ensure that all eight ADs have the same Gd content.
The LS is also produced in 4-ton batches and then stored in a 200-ton storage pool with internal circulation. The MO is stored in a different 200-ton pool. The Gd-LS has 87.7\% carbon content by weight, 12.1\% hydrogen, and 0.103\% Gd. The density of Gd-LS, LS, and MO is 0.860, 0.859, and 0.851 g/ml, respectively.

\par
To reduce systematic uncertainties and to reduce contamination from dust, all ADs are assembled in pairs in a large ISO 7 (class 10,000) clean room in the SAB.   Each pair is filled over a short interval of time so that detector pairs are matched in Gd-LS quality and characteristics. Gd-LS, LS, and mineral oil filling occurs underground in the LS Hall.  All three liquids are filled concurrently to maintain a uniform liquid level across all three volumes.  MO and LS are pumped directly from the large storage reservoirs, and their mass is measured by Coriolis flow meters.  Gd-LS is first transferred equally from five separate Gd-LS storage tanks to a 20-ton ISO tank lined with Teflon and instrumented with load cells before being pumped into the AD.  The AD target mass can be determined to O($0.1\%$) by weighing the ISO tank before and after the filling.  All ADs are filled with the same filling system.

\par
All detector components passed low background testing. The stainless steel of the SSV was specially made in one batch with low-radioactivity iron ore.  PMTs with low background glass were procured.  Chemicals used for synthesizing the Gd-LS, GdCl$_3$, TMHA, and PPO were purified to get rid of some radioactive contaminants. The liquid scintillators are covered with a continuous flow of dry nitrogen gas to prevent oxidation while in storage and within an AD.

\subsection{Muon system}
\par

\begin{figure}[htb]
\centering
\includegraphics[width=7.0cm]{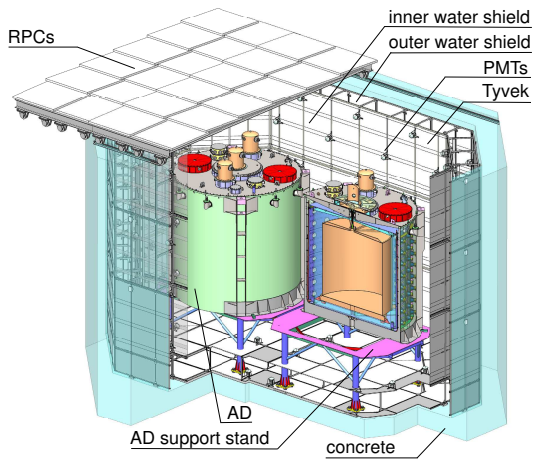}
\caption{Schematic for the Daya Bay Near Hall (EH1) including the ADs, water shields, and RPCs.}
\label{fig:muon}
\end{figure}

The muon detector consists of a RPC tracking device and an active water shield.  The water shield consists of two optically separated regions known as the inner (IWS) and outer (OWS) water shields. Each region operates as an independent water Cherenkov detector instrumented with PMTs. The water shield as a whole has multiple purposes. It detects muons that can produce spallation neutrons or other cosmogenic backgrounds in the ADs. The pool also moderates neutrons and attenuates gamma rays produced in the rock and other structural materials in and around the experimental halls. The water pool is designed so that there is at least 2.5 m of water surrounding each AD in every direction, as seen in Fig.~\ref{fig:muon}. The near site water pools each contain 1200 t of purified water produced at 18 M$\Omega$-cm while the far water pool contains 1950 t of purified water. The PMTs are distributed between the inner and outer zones that are optically divided by Tyvek sheets.  Each pool is outfitted with a light-tight cover with dry-nitrogen flowing underneath.

Each water pool is covered with an array of RPC modules~\cite{rpc}.  The 2 m x 2 m modules are deployed in an overlapping pattern (that minimizes dead areas) on a steel frame mounted on rails, so that the assembly can be retracted to provide access to the water pool.  There are four layers of bare RPCs inside each module, with one layer of
readout strips associated with each layer of bare RPCs.  The strips have a "zigzag" design with an effective width of 25 cm, and are stacked in alternating orientations providing a spacial resolution of $\sim$8 cm.

\subsection{Electronics}

\begin{figure}[htb]
\centering
\includegraphics[width=7cm]{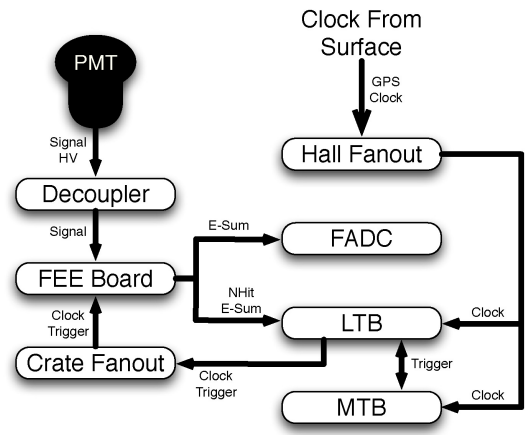}
\caption{Block diagram outlining the PMT readout electronics. }
\label{fig:PMT_Readout}
\end{figure}

Each detector unit (AD, IWS, OWS, and RPC) is read out by a single VME crate.  There are two types of readout crates, PMT readout and RPC readout.   All PMT readout crates are physically identical, differing only in the number of instrumented readout channels.  Each near site has four PMT readout crates and one RPC readout crate, while the far site has six PMT readout crates and one RPC readout crate.  The PMT readout is documented elsewhere~\cite{FEE_IEEE} and is shown schematically in Fig.~\ref{fig:PMT_Readout}.  In brief, the front-end electronics board (FEE) receives raw signals from up to sixteen PMTs, sums the charge among all input channels, identifies over-threshold channels, records timing information on over-threshold channels, and measures the charge of each over-threshold pulse.  The FEE in turn sends the number of channels over threshold and the integrated charge to the trigger system as well as the FADC board.  The FADC samples and records the unshaped energy sum at 1 GHz.  When a trigger is issued, the FEE reads out the charge and timing information for each over-threshold channel, as well as the average ADC value over a 100 ns time-window immediately preceding the over-threshold condition (preADC).

The RPC readout consists of 32-channel front-end cards (FECs) mounted on the detector modules. Each FEC reads out one RPC detector module. The RPC trigger module (RTM) and the RPC readout module (ROM) sit in a VME crate in the electronics room.  Transceiver modules in custom crates mounted on the RPC detector frame relay trigger and timing information between the RTM and FECs, while also transmitting hit information to the RTM and ROM.  The RPC readout is independent of the PMT readout and consists of a digital hit map of the over-threshold channels along with a GPS time-stamp for the trigger.  Additional details can be found in~\cite{RPC_IEEE}.

\subsection{Triggers}
\par

Triggers are primarily created internally within each VME crate, although the system is also designed to accept external trigger signals.  Each PMT readout crate contains a local trigger board (LTB) that receives the number of over-threshold channels (Nhit) as well as the summed charge (E-Sum) from each FEE~\cite{trigger}.  The Nhit and E-Sum data form the basis for two physical triggers, the multiplicity trigger and energy-sum trigger.   In addition to the physical triggers, the LTB can also generate two internal triggers, a periodic trigger and a software trigger. Additionally, the LTB can receive and process two external trigger requests, the calibration trigger request and a cross trigger request. Cross triggers allow one detector subsystem to trigger other detector subsystems. Each type of trigger can be enabled or disabled via VME interface. The enabled triggers are ORed to generate a final trigger signal (local trigger) that is distributed to the FEEs by front-panel cables.  This design is flexible and provides various options to study and increase the overall trigger efficiency. The RPC sub-detector uses custom front-end electronics that employ a self-triggering scheme.  A master trigger board (MTB) coordinates cross triggers, allowing a sub-detector to be triggered by another sub-detector.  Calibration trigger requests are used to monitor detector performance and come into the system through the MTB.  The MTB broadcasts this trigger to the appropriate LTB as a cross-trigger command.
To avoid an occurrence of overflow in either the LTB data-buffer or the FEE data-buffer, the LTB can check if either is nearly full. If so, the local trigger and the corresponding data package can be blocked. The blocked trigger number is recorded and read out for calculating the dead time of the trigger-system offline.

\subsection{Data acquisition}
\par
The data acquisition (DAQ) architecture is designed as a multi-level system using embedded Linux, advanced commercial computer and distributed network technology, and is modeled after the BESIII and ATLAS DAQ systems~\cite{daq}. The readout DAQ operates concurrently within each VME crate from a MVME 5500 single board computer running TimeSys (a realtime LINUX based OS).  Event readout begins with an interrupt from the trigger module that initiates read out of data fragments from all modules residing in the crate using chained block transfer (CBLT) mode.  Fragments are concatenated into an event ordered in time to create a data stream for that VME crate.  The front-end VME system transmits the data to the back-end DAQ that merges and sorts the events from the various incoming streams  by trigger time-stamp.  The DAQ system also provides an interactive interface for the electronics, trigger and calibration systems.  Run control is flexible and configurable, allowing global operation of all detector systems or operation of sub-sets of detectors whenever debugging or commissioning is required.

\subsection{Data stability and selection}
\par
Stable data taking in EH1 began September~23,~2011.  Table~\ref{tbl:data} summarizes the experimental livetime from September~23,~2011 to December~23,~2011.  The fraction of the physics-data-taking (non-calibration and non-diagnostic) time in total calendar time (2194.3 hours) was 86.8\%. Furthermore, the data quality is good with 1684.2 hours of data deemed suitable for physics analysis. We excluded 221.0 hours of data from physics analysis, including 203.2 hours of systematic studies, 10.2 hours due to some coherent noise pickup in the electronics, and 7.6 hours due to electronics, high voltage, or DAQ problems.  The systematic studies primarily occurred between December~10 and December~18, 2011 when the HV for certain PMTs were lowered to investigate 'flasher' events (see Section 5.4).  The DAQ dead time due to full data buffers was determined with dedicated scalers to be less than 0.0025\%  for both AD1 and AD2.

\begin{table}[!ht]
   \centering
\begin{tabular}{ l l }
\hline\hline
   Total calendar time (hr) & 2194.3\\
   Total DAQ time (hr) & 2092.5 \\
   Physics DAQ time (hr) & 1905.2 \\
   Good run time (hr) & 1684.2 \\   \hline
\end{tabular}
\caption{Summary of experimental livetime.\label{tbl:data}}
\end{table}

\subsection{Offline processing}
During data taking, the onsite experimental raw data are transferred to the computing centers located at the Institute for High Energy Physics (IHEP) in Beijing as well as Lawrence Berkeley National Laboratory (LBNL) in California in real time. The multi-flow concurrent transmission mode is adopted to improve the data transmission bandwidth. As soon as the data reach the destinations, a {\em keep-up} data processing takes place. To help assess and monitor data quality, distributions of selected variables are automatically plotted and published on the web.  After checking the data quality, the offline data processing reconstructs and tags the data for physics analysis.

\par
The Daya Bay offline software (known as NuWa) has been developed using Gaudi~\cite{gaudi} as the underlying software framework in order to provide the full functionality required by simulation, reconstruction and physics analysis. NuWa employs Gaudi's event data service as the data manager. Raw data, as well as other offline data objects, can be accessed from the Transient Event Store (TES). According to the requirement from the prompt-delayed coincidence analysis, another specific Archive Event Store (AES) has also been implemented to provide the function of look-back in time. All the data objects in both TES and AES can be written into or read back from ROOT~\cite{root} files through various Gaudi converters. In this framework, reconstruction algorithms have been developed to construct the energy and the vertex of the antineutrino event from the charge pattern of the PMTs. The detector-related parameters and calibration constants needed by the reconstruction are stored in an offline central database with a number of mirror sites located at different institutes. The algorithms can access the contents in the database via an interface software package called DBI.

\par
The Daya Bay simulation is based on GEANT4~\cite{GEANT4} with certain critical features validated against external data, where available, or other simulation packages such as MCNPX~\cite{MCNPX}, FLUKA~\cite{FLUKA}, GCALOR~\cite{GCALOR} and GEANT3.  Features such as the gamma spectrum from a neutron capture on Gadolinium, the cosmic muon flux, and specific decay chains have been custom built. Tracking of optical photons is used during high precision simulations, and a comprehensive electronic readout simulation has been implemented. The Monte Carlo (MC) simulation is tuned to match observed detector response.  To simulate time correlations of unrelated events, different categories of simulated events are mixed in the output file.

\section{Low level performance}
\par

\subsection{ADC calibration}
\par
\begin{figure}[htb]
\centering
\includegraphics[width=7cm]{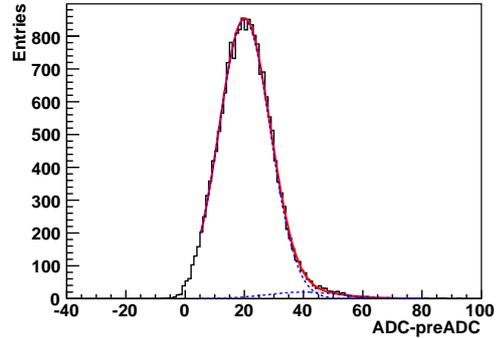}
\caption{Fit to a single PMT SPE distribution using the function specified in equation~\ref{eq:GainSPE} after subtracting the baseline. Dashed curves show the SPE and double PE components.
The units are in raw ADC counts.}
\label{fig:GainSPE}
\end{figure}

Each readout channel is calibrated with data collected in dedicated low-intensity LED runs.  The calibration is verified using singles collected continuously during regular data running.   In either case, samples dominated by single photo-electron (SPE) hits are selected and fit after baseline subtraction using a function as shown in Eq.~\ref{eq:GainSPE}.

\begin{equation}
\hspace{-.15in}
S(x) = \displaystyle\sum\limits_{n} \frac{\mu^n e^{-\mu}}{n!} \frac{1}{\sigma_1 \sqrt{2n\pi}}exp(-\frac{(x-nQ_1)^2}{2n\sigma_1^2})\,,
\label{eq:GainSPE}
\end{equation}
which is a convolution of Poisson distributions with a Gaussian. Care is taken to ensure that SPE hits are dominant in the fit by requiring the PMT occupancy to be less than 0.13 and that the preceding analog baseline is stable. An example of the fitting is shown in Fig.~\ref{fig:GainSPE}.

\par
\begin{figure}[htb]
\centering
\includegraphics[width=7cm]{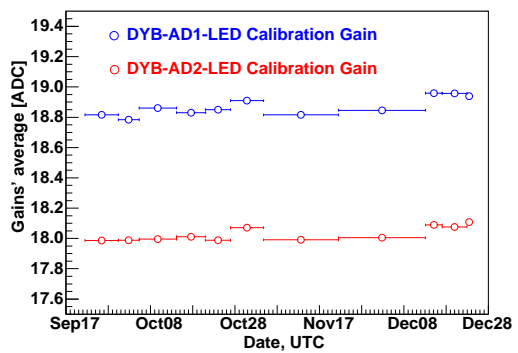}
\caption{The stability of the peak in the SPE distributions averaged across each AD over a period of three months.}
\label{fig:GainStability}
\end{figure}
\begin{figure}[htb]
\centering
\includegraphics[width=7cm]{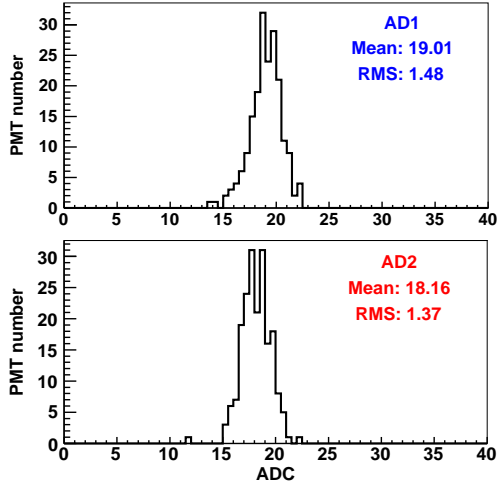}
\caption{Distribution of SPE peaks for AD1 and AD2 for data taken in late December.
\label{fig:AD12GainDist}}
\end{figure}

The fit values are stored in a database for use during data processing.  We find a systematic difference in the average ADC count per SPE between AD1 and AD2.  We also observe a slight drift in the calibration as a function of time; however, all channels appear to drift together and the difference between AD1 and AD2 is stable, as shown in Fig.~\ref{fig:GainStability}.  The distribution of calibration constants, ADC counts per photo-electron (p.e.), for  AD1 and AD2 is shown in Fig.~\ref{fig:AD12GainDist} for one of the calibration periods.   The channel to channel variation is within expectations.

\subsection{PMT singles rate}

\par
\begin{figure}[htb]
\centering
\includegraphics[width=7cm]{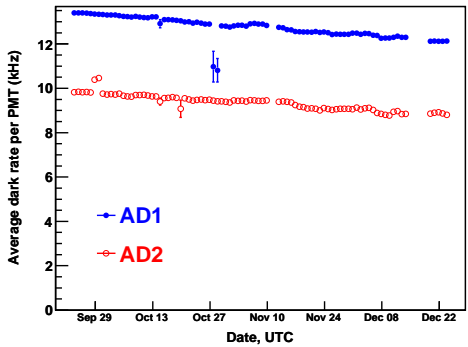}
\caption{Distribution of PMT dark rates for AD1 and AD2.}
\label{fig:DarkRate}
\end{figure}

The PMT singles rates are studied using periodic triggers.  The observed dark rates for AD1 and AD2 are relatively stable and slowly decreasing (by about 10\%) over several months, as shown in Fig.~\ref{fig:DarkRate}.  With the channel thresholds set near 1/4 p.e., the average dark rates of all PMTs are about 10 kHz for AD1 and about 12 kHz for AD2.

\subsection{Trigger}

\par
The performance of the AD trigger system was probed by placing radioactive sources and LEDs at different locations within the target volume. Positrons from a $^{68}$Ge source were used as a proxy for IBD prompt events. The threshold of one of the two redundant trigger modes (Nhit and Esum) is lowered to test the trigger response of the other mode. In order to measure the trigger response to weak LED light, an external trigger, synchronized to the LED driver, is used. Both methods gave consistent results.
\begin{figure}[htb]
  \centering
    \includegraphics[width=7cm]{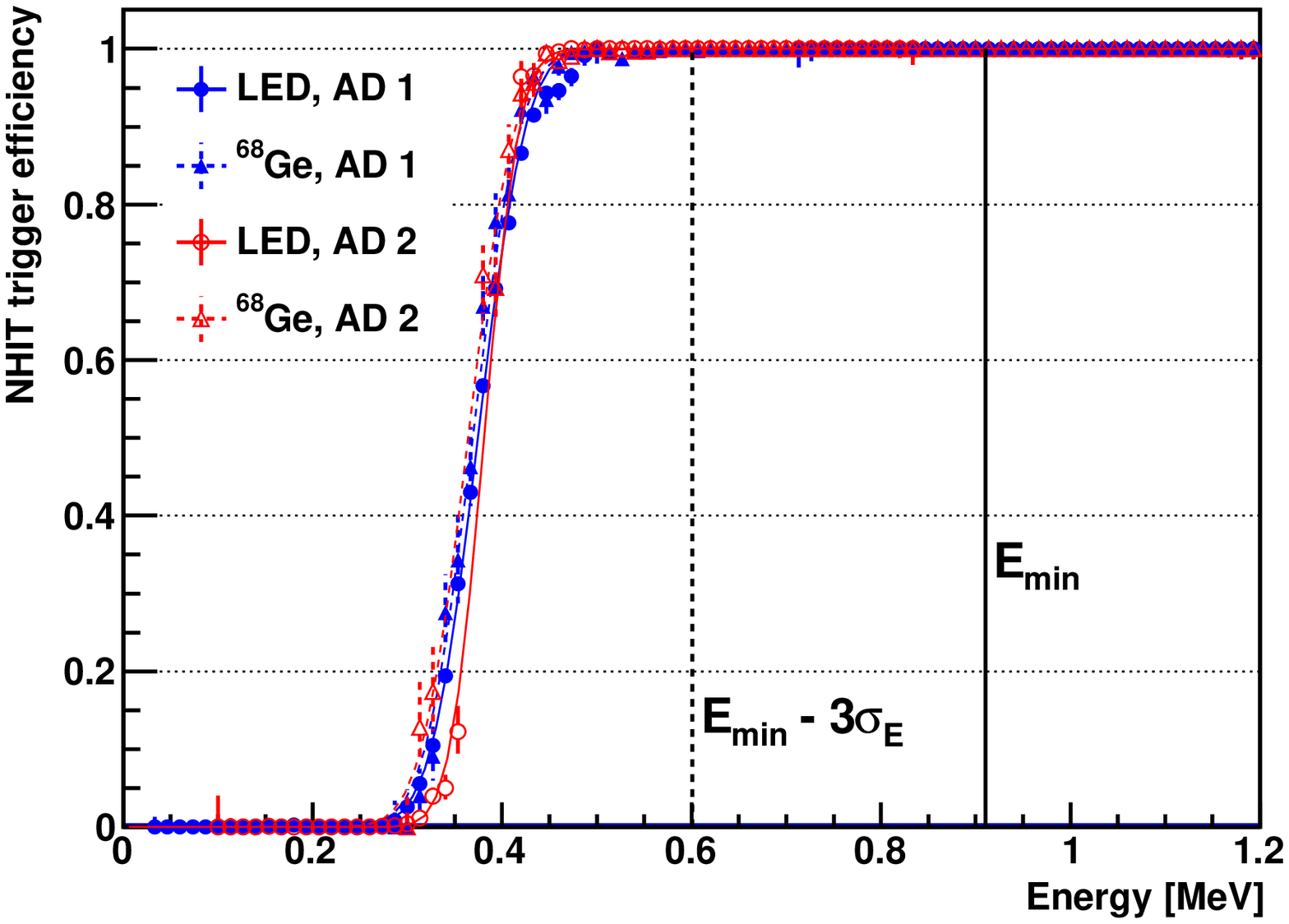}
    \includegraphics[width=7cm]{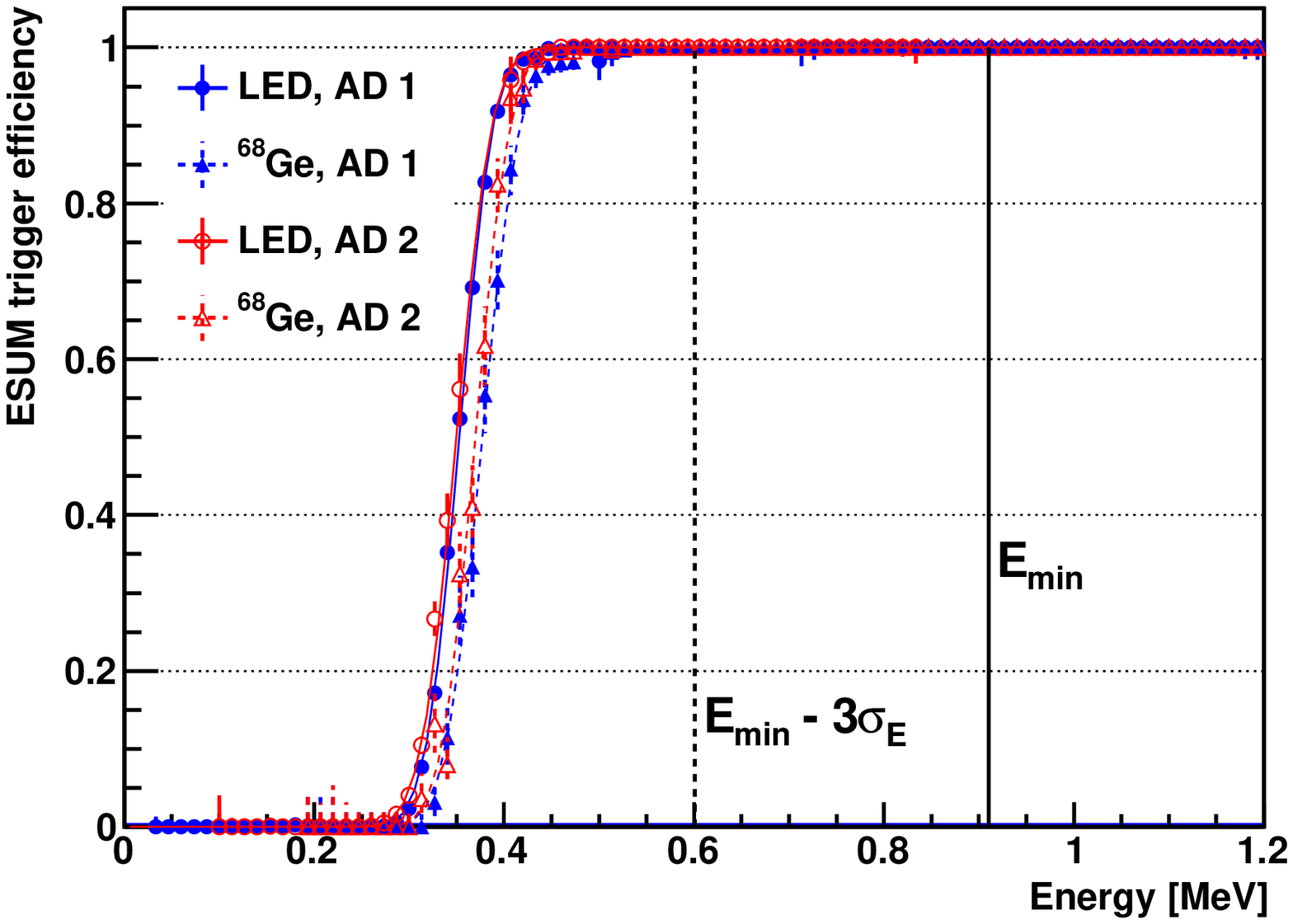}
  \caption{Trigger efficiency as a function of reconstructed energy at the edge of the AD target volume ($r=120\,\rm{cm}$, $z=135\,\rm{cm}$). The top figure is for Nhit triggers while the bottom figure is for E-Sum triggers.  Triangles represent efficiency measurements from $^{68}$Ge source data. Circles result from LED scans. The colored lines show best fits based on error functions. The black lines indicate minimum reconstructed energies $E_{\rm min}$ of IBD positrons.  \label{fig:trigeff}}
\end{figure}

\par
Figure \ref{fig:trigeff} shows the measured trigger efficiency as a function of reconstructed energy at the edge of the AD target volume. The energy-equivalent of the trigger thresholds, determined by fitting error functions against the data, was measured at $E_{\rm{th}} \sim 0.37\,\rm{MeV}$ for both trigger modes.

\par
The energy reconstruction will be described in later sections and does not correct for the non-linear response of the scintillator. Thus, a positron annihilation is reconstructed at a mean energy of $E_{\rm min} \sim 0.9\,\rm{MeV}$ with an energy resolution $\sigma_{E} \sim 0.1\,\rm{MeV}$. At energies $E_{\rm min}-3\sigma_{E}$, the trigger efficiency is still $\epsilon = 1^{+0}_{-0.002}$ throughout the scintillating volume, thus ensuring negligible trigger inefficiency.

\section{Comparison of rates and energy spectra}
\par

\subsection{Calibration sources\label{sec:source}}
\par
The ADs are calibrated periodically using the automated calibration units on the lid of each detector. The calibration program includes routine weekly calibration, as well as dedicated detector studies. During commissioning, the detector response is studied extensively with high statistics using all three sources in each of the three ACUs, including a fine-grain scan along each vertical axis. During weekly calibration, the combined Am-$^{13}$C/$^{60}$Co sources are lowered separately to five vertical locations for all three ACUs. Data are collected for five minutes at each position. This is followed by a $^{68}$Ge source deployed to five vertical locations for ACU-A, each for five minutes of data taking.

\par
\begin{figure}[htb]
\centering
\includegraphics[width=7cm]{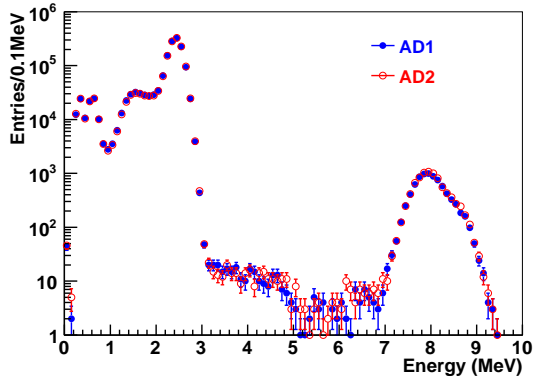}
\caption{The energy spectra observed using the Am-$^{13}$C/$^{60}$Co sources at the center of AD1 and AD2 are overlaid.}
\label{fig:cospec}
\end{figure}

The energy spectra observed using the Am-$^{13}$C/$^{60}$Co sources at the center of AD1 and AD2 are shown in Fig.~\ref{fig:cospec}. Background spectra are taken from physics runs (without source deployment) and subtracted by normalizing the background run time to the calibration run time. Backgrounds can be reduced by applying a vertex cut around the source position.
Besides the neutron peak at 8.05 MeV and the cobalt peak at 2.51 MeV, there is a single gamma shoulder at about 1.5 MeV and two peaks at lower energies. The spectra for AD1 and AD2 are in good agreement. The source activities measured by fitting the spectra are similar for the two ADs.

\par

\begin{figure}[htb]
\centering
\includegraphics[width=7cm]{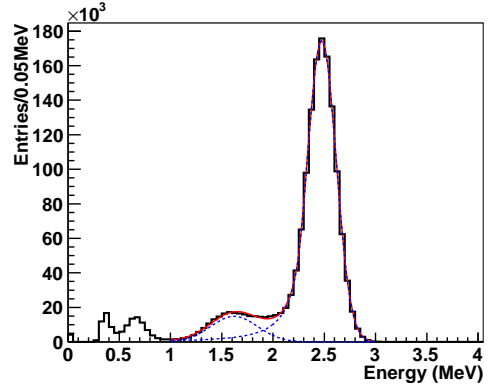}
\caption{The cobalt spectrum fit using a Crystal Ball function plus a Gaussian. }
\label{fig:cofit}
\end{figure}

The cobalt peak is fit with a Crystal Ball function~\cite{cbfunction} plus a Gaussian (see Fig.~\ref{fig:cofit}).
The Crystal Ball function is given by
\begin{equation}
 \hspace{-0.5cm} f(x;\alpha,n,\bar x,\sigma) = N \cdot \left\{
  \begin{array}{ll}
    \exp(- \frac{(x -\bar x)^2}{2 \sigma^2})      & \mbox{for } \frac{x - \bar x}{\sigma} > -\alpha \\
    A (B - \frac{x - \bar x}{\sigma})^{-n}  & \mbox{for } \frac{x - \bar x}{\sigma} \leq -\alpha
  \end{array}
  \right.
\end{equation}
where:
\begin{eqnarray*}
  A & = & \left(\frac{n}{\left| \alpha \right|}\right)^n \cdot \exp\left(- \frac {\left| \alpha \right|^2}{2}\right) \\
  B & = & \frac{n}{\left| \alpha \right|} - \left| \alpha \right|
\end{eqnarray*}
$N$ is a normalization factor and $\alpha$, $n$, $\bar x$ and $\sigma$
are parameters which are fit with the data. The Crystal Ball function is chosen to separate the Gaussian response of the detector from the low energy tail, thus avoiding bias from the asymmetry of the spectrum. The small Gaussian component models the absorption of a gamma ray from the $^{60}$Co decay by the source packaging material.

\par
\begin{figure}[htb]
\centering
\includegraphics[width=7cm]{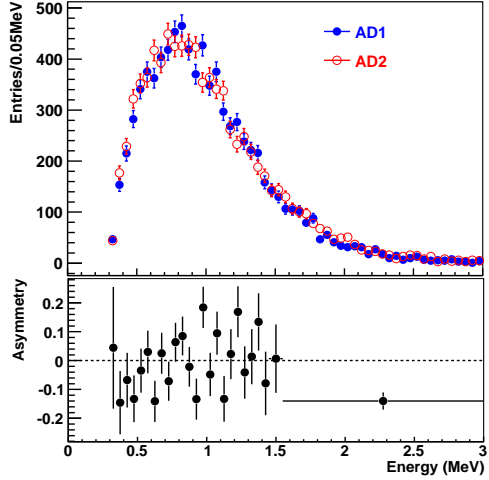}
\caption{The prompt signal from the proton recoil of Am-$^{13}$C neutron. }
\label{fig:amcprompt}
\end{figure}

The Am-$^{13}$C source produces neutrons with an average kinetic energy of 3.5 MeV.  Although the rate is low, neutrons from the Am-$^{13}$C source can be cleanly selected by requiring a coincidence between the prompt proton recoil signal and the delayed neutron signal. No background subtraction or vertex cut selection is required. The prompt signal from the proton recoil is shown in Fig.~\ref{fig:amcprompt}. The ADs have matching spectra, indicating that the energy response and quenching of the Gd-LS is the same in both.  To better quantify the difference in the observed spectra, Fig.~\ref{fig:amcprompt} plots a variable called {\it Asymmetry} defined as
\begin{equation}
\label{eq:asymmetry}
{\rm Asymmetry}=\frac{N_{\rm AD1}-N_{\rm AD2}}{(N_{\rm AD1}+N_{\rm AD2})/2}
\end{equation}
where $N_{\rm AD1,2}$ is the bin content for AD1 or 2, respectively.

\par
\begin{figure}[htb]
\centering
\includegraphics[width=7cm]{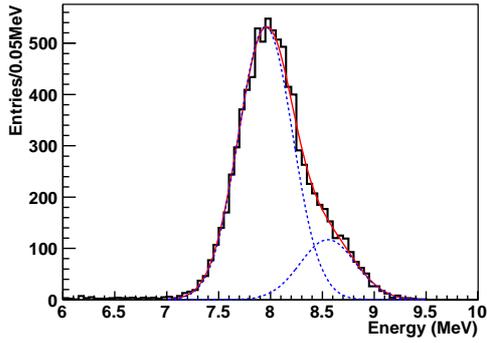}
\caption{The neutron spectrum of the Am-$^{13}$C source at the detector center, fitted with two constrained Gaussians. }
\label{fig:neutronfit}
\end{figure}

When a neutron is captured by Gd, it has  a 0.1848 probability of being captured by $^{155}$Gd releasing gamma rays with a total energy of 8.536 MeV, and a 0.8151 probability of being captured by $^{157}$Gd, releasing 7.937 MeV.  Contributions from other Gd isotopes can be ignored.  For an event at the center of the detector, the neutron peak can be fit with two Gaussians with their relative area constrained by their neutron capture probabilities and the relative widths constrained by $1/\sqrt{E}$ energy resolution, as shown in Fig.~\ref{fig:neutronfit}. The free parameter for the peak is taken as the average of the two peaks weighted by their neutron capture probabilities. For an event close to the edge of the IAV, two Crystal Ball functions are used for fitting, with similar constraints as above.

\par
\begin{figure}[htb]
\centering
\includegraphics[width=7cm]{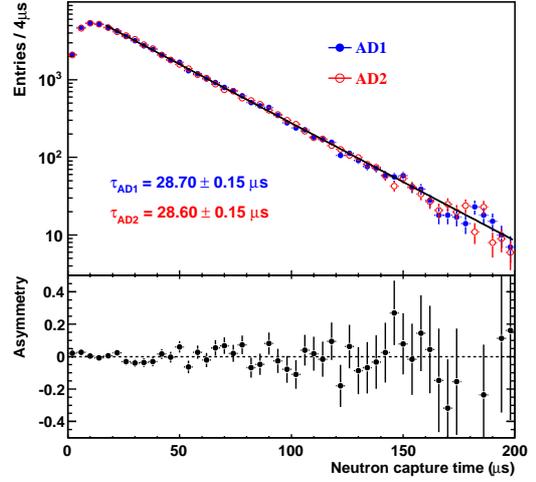}
\caption{The neutron capture time on Gd from the Am-$^{13}$C source at the detector center. }
\label{fig:amctime}
\end{figure}

The neutron capture time on Gd with the Am-$^{13}$C source at the detector center is shown in Fig.~\ref{fig:amctime}. Since the capture time is directly related to the Gd concentration, Fig.~\ref{fig:amctime} provides a measure of the identicalness of the Gd concentration in AD1 and AD2. The measured capture times are $28.70\pm0.15~\mu s$ and $28.60\pm0.15~\mu s$ for AD1 and AD2, respectively.

\par
\begin{figure}[htb]
\centering
\includegraphics[width=7cm]{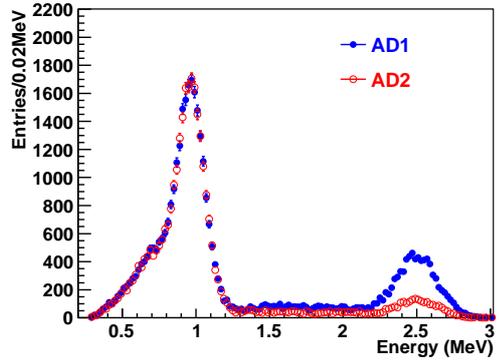}
\caption{The energy spectrum of the $^{68}$Ge source.}
\label{fig:gespec}
\end{figure}
\begin{figure}[htb]
\centering
\includegraphics[width=7cm]{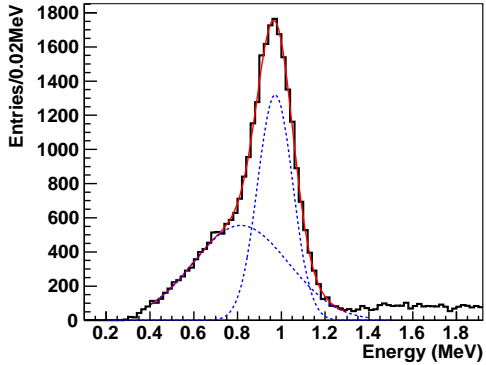}
\caption{The $^{68}$Ge spectrum fitted with two Gaussians.}
\label{fig:gefit}
\end{figure}

Fig.~\ref{fig:gespec} displays the observed spectrum from the $^{68}$Ge source deployed at the center of the detector.  $^{60}$Co contamination in the source is responsible for the peak at 2.5 MeV and the tail, which is the shoulder in Fig.~\ref{fig:cofit}, extending to lower energy. While undesirable, the impact of the $^{60}$Co contamination on the fitting of the $^{68}$Ge peak is negligible. The $^{68}$Ge activities and spectra for the two ADs are similar while the contaminations are different. The peak is fit with two Gaussians, one describing the full energy deposit in the detector and the other describing the energy loss in the source enclosure, as shown in Fig.~\ref{fig:gefit}.

\subsection{AD energy response}

\par
The AD energy is reconstructed based on the total charge of an event, which is the charge sum of all 192 8-inch PMTs. The visible energy $E_{\rm vis}$ is defined as the total charge divided by an energy calibration constant.

\begin{equation}   E_{vis} = Q_{tot}/C  \end{equation}

\noindent We use two independent energy reconstructions, as described in Sec.~\ref{sec:erec}.
One approach is to determine the calibration constant by constraining the energy peak of the $^{60}$Co at the detector center to 2.506 MeV. This energy calibration is done weekly, at the same frequency as the PMT gain calibration. A second approach using spallation neutrons captured on Gd is also used to derive and monitor the energy calibration constant. For this method, the calibration constants are updated more frequently
during regular physics data runs.

\par
\begin{figure}[htb]
\centering
\includegraphics[width=7cm]{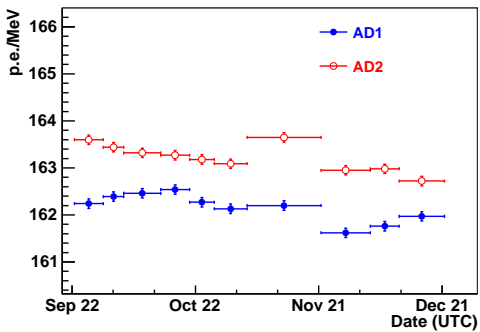}
\caption{The time stability of the energy calibration constants. The drift relative to the first data point is shown in the bottom panel.}
\label{fig:tstable}
\end{figure}

For the $^{60}$Co calibration, the energy constant is about 162 p.e./MeV for AD1 and about 163 p.e./MeV for AD2. The time dependence is shown in Fig.~\ref{fig:tstable}.  For each AD, the values vary within a narrow band with a width of 0.4\%. The time dependence is also monitored using the visible energy from the $^{60}$Co source at five vertical locations for all three ACUs, and the $^{68}$Ge source from ACU-A deployed at five vertical locations. The time dependence of the reconstructed energy is largely corrected by the weekly calibration.  The systematic uncertainty is determined by taking an average of the differences between consecutive calibrations (p.e. per MeV), which is 0.2\%.

\par
\begin{figure}[htb]
\centering
\includegraphics[width=7cm]{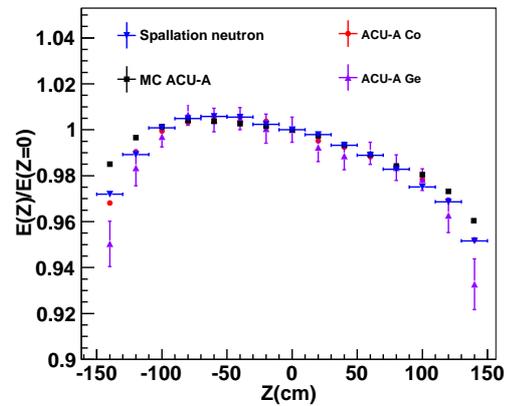}
\caption{The AD energy response as a function of {\em z} for Monte Carlo data, spallation neutron data, and ACU-A sources.
\label{fig:cozscan}}
\end{figure}

For a perfect detector with reflective panels at the top and bottom, the energy response should be uniform along the vertical axis.  A z-scan along the central axis using the $^{60}$Co source is shown in Fig.~\ref{fig:cozscan}.  A z-position of -150 cm corresponds to the bottom of the 3-m AV while +150 cm is the top.
The detector response shows good uniformity along z direction; however, two features exist.  One is a decrease in response at the top and bottom due to absorption in the reinforcing ribs of the AVs and the non-ideal reflectivity of the reflectors. The other is an asymmetry introduced by the reflector locations. The PMT array is centered on the bulk of liquid scintillator; however, the 3-m and 4-m AVs have cone-shaped tops. Therefore, the gap between the top ring of PMTs and the top reflector is larger than the gap between the bottom ring of PMTs and the bottom reflector by 77 mm. This is a geometric effect that is well reproduced in Monte Carlo simulations. If we center the PMT array between the reflectors, the simulated detector response becomes symmetric along the z axis. The same non-uniformity is also verified by $^{68}$Ge source and spallation neutrons, as shown in Fig.~\ref{fig:cozscan}. For the spallation neutrons, only those in a small cylinder around the central axis are selected for comparing with ACU-A data.

\begin{figure}[htb]
\centering
\includegraphics[width=7cm]{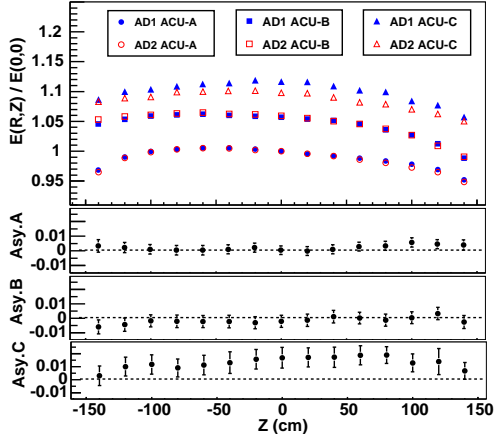}
\caption{Visible energy in {\em z} with respect to the detector center for $^{60}$Co sources from each ACU is shown at top.  The corresponding Asymmetry parameter is shown for ACU-A (Asy. A), ACU-B (Asy. B) and ACU-C (Asy. C) in the bottom panels. }
\label{fig:cozscan3}
\end{figure}

The z-scan with the $^{60}$Co sources from all three ACUs is shown in Fig.~\ref{fig:cozscan3}. The energy response for ACU-B and ACU-C exhibit similar features as ACU-A described above, but with more visible energy. This is also a geometric effect caused by an increase in average acceptance as the source approaches the PMTs. It is well reproduced in Monte Carlo, and is linear in $r^2$, where $r$ is the radial position of the source. The differences between the ADs are small for all three ACU scans. Analogous to Eq.~\ref{eq:asymmetry}, we define the asymmetry between the fitted energy peaks ($\mu$)
\begin{equation}\label{eqn:asymE}
	{\rm Asymmetry} = \frac{\mu_{\rm AD1} - \mu_{\rm AD2}}
	{(\mu_{\rm AD1} + \mu_{\rm AD2})/2} \ \ \ .
\end{equation}
The Asymmetry values lie within a narrow band with a width of 0.3\%  for ACU-A, 0.4\% for ACU-B, and 2\% for ACU-C. The ACU-C scan is in the gamma catcher, and may suffer more from edge effects, such as acrylic properties and PMT-to-PMT efficiency variations.

\par
\begin{figure}[htb]
\centering
\includegraphics[width=7cm]{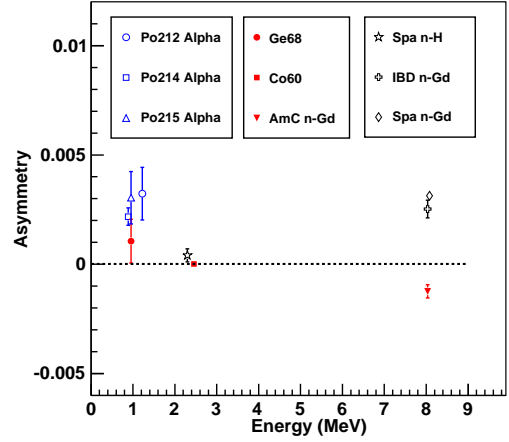}
\caption{Asymmetry of energy nonlinearity for two ADs. See text for details.}
\label{fig:nonlin}
\end{figure}

Detector response (observed energy versus deposited energy) is not linear in energy, and the effect differs with particle species.  This nonlinearity is due to the quenching effects of the liquid scintillator, Cherenkov light emission and subsequent absorption and re-emission in the liquid scintillator, PMT dark noise, PMT or electronic non-linearity, etc. However, the effects are the same in each AD as shown in Fig.~\ref{fig:nonlin}.  In this figure, the three radioactive sources are deployed at the detector center, and all alphas, spallation neutrons, and IBD neutrons are  distributed uniformly throughout the detector. The alphas result from daughter decays from residual contamination of ${}^{238}$U, ${}^{232}$Th and ${}^{227}$Ac. Clean samples of alphas from ${}^{214}$Po, ${}^{212}$Po and ${}^{215}$Po decays are selected using the $\beta-\alpha$ or $\alpha-\alpha$ time correlations. The energy is vertex-corrected, as described in Sec.~\ref{sec:erec}.

\subsection{Energy reconstruction\label{sec:erec}}

\par
Given the importance of minimizing the energy scale uncertainties in the Daya Bay analysis, two independent energy reconstructions are adopted to cross check each other. One is based on source calibration data and the other uses spallation neutrons.
Both apply a vertex correction to the visible energy.  Since minimizing the event selection criteria is important to maintaining a small systematic error, vertex reconstruction will not be used to select IBD events. However, vertex reconstruction is necessary to correct the position dependent energy non-uniformity. It is also valuable for many other studies.

\subsubsection{Energy reconstruction based on sources\label{sec:er1}}
\par
The vertex reconstruction is based on center-of-charge (COC), defined as the charge-weighted-mean of the coordinates of all PMTs. The mapping from COC to vertex is done by analytic corrections. The correction formula is inspired from Monte Carlo simulations and is almost linear in COC but with small corrections. The reconstructed vertex is
\begin{eqnarray}
\label{eqn:vtxrcorr}
\hspace{-0.3cm} r &=& c_1 \times R_{\rm COC} - c_2 \times R^2_{\rm COC} \,,\\
\label{eqn:vtxzcorr}
\hspace{-0.3cm} z &=& (Z_{\rm COC}-c_3 \times Z^3_{\rm COC})\times (c_4-c_5 \times R_{\rm COC})\,,
\end{eqnarray}
where $c_1-c_5$  are coefficients that are found by calibrating the COC to the true locations of the cobalt sources with data.  $R_{\rm COC}$ and $Z_{\rm COC}$ are the COC coordinates.

\par
To correct the vertex-dependent energy non-uniformity, cobalt source z-scan data is used to model the corrections in terms of the radius $r$ and $z$. The reconstructed energy is
\begin{equation}
\label{eqn:erec}
E^{Co}_{\rm rec} = C^{Co}_n \times  E_{\rm vis}/(C^{Co}_r(r)\times C^{Co}_z(z)) \,,
\end{equation}
where $E_{\rm vis}$ is defined in section 5.2,
$C^{Co}_r(r)$ is a linear function of $r^2$, $C^{Co}_z(z)$ is a 3rd-order polynomial, and $C^{Co}_n$ is a constant determined by the ratio of the neutron peak of the Am-$^{13}$C source to the cobalt peak, both at the detector center. $C^{Co}_n$ rescales the corrected energy from the cobalt energy scale to the neutron energy scale. Both AD1 and AD2 data are used to extract the coefficients in Eq.~\ref{eqn:vtxrcorr}-\ref{eqn:erec}.
All AD events will be reconstructed to the neutron energy scale first.

\par
\begin{figure}[htb]
\centering
\includegraphics[width=7cm]{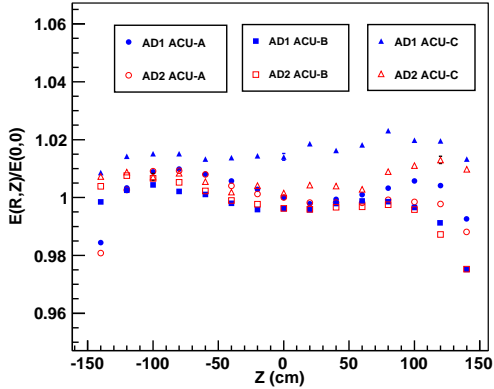}
\caption{Reconstructed energy in z with respect to the detector center for $^{60}$Co sources from each ACU, after the vertex-dependent correction.}
\label{fig:reczscan}
\end{figure}

As displayed in Fig.~\ref{fig:reczscan}, after the vertex-dependent correction, the non-uniformity of visible energy shown in Fig.~\ref{fig:cozscan3} has been minimized. All data points for each ACU in both ADs now lie within a band from 0.98 to 1.02, relative to the cobalt energy response at the detector center.  Since we did not apply detector-dependent corrections, and the vertex-dependent correction factor $C^{Co}_r(r)\times C^{Co}_z(z)$ is at most 10\% within the target and gamma catcher volume, the reconstructed energy has nearly the same relative non-uniformity
between the ADs as the visible energy.

\subsubsection{Energy reconstruction based on spallation neutrons\label{sec:er2}}

\begin{figure}[htb]
\centering
\includegraphics[width=7cm]{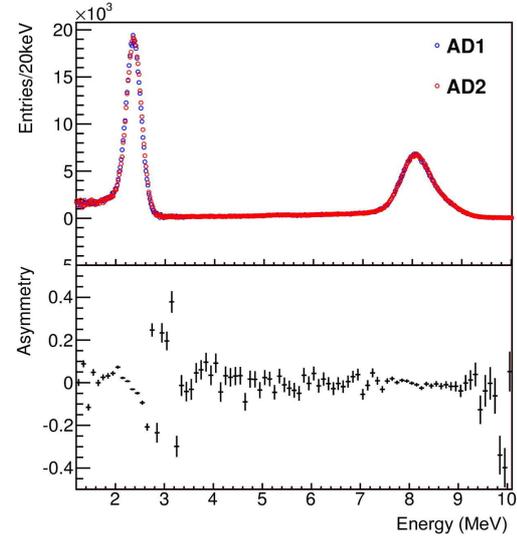}
\caption{Spallation neutron energy spectrum and Asymmetry between ADs.}
\label{fig:spapeak}
\end{figure}

More than 8,000 spallation neutrons per detector per day are observed at the near site. They can be cleanly selected by searching in a time window from 20 to 200 $\mu$s after a muon passes through an AD. The lower limit is required to ensure that the baseline of the electronics has recovered after a large muon signal.  Capture on both Gd and H can be seen in the spallation neutron spectrum in Fig.~\ref{fig:spapeak}. The small Asymmetry shows that the spectra for the two ADs are almost the same.
\par
The vertex position is reconstructed by comparing the number of photo-electrons observed in each PMT with templates produced through MC simulation. Templates are produced for 20 bins in the $r$ direction, 20 bins in the $z$ direction, and 24 bins in the $\phi$ direction. For each grid point where the templates are produced, we compute the $\chi^2$ values defined as:
\begin{equation}
\chi^2  = \sum_i^{\rm PMTs}\left[  -2 \ln \frac{P(N^{\rm obs}_i,N^{\rm exp}_i(r,z,\phi))}{P(N^{\rm obs}_i,N^{\rm obs}_i)}\right],
\end{equation}
where  $P(n,\mu) = \mu^ne^{-\mu}/(n!)$ is the probability of finding $n$ photo-electrons when the mean value is $\mu$ assuming Poisson statistics, $N^{\rm obs}_i$ is the observed number of photo-electrons and $N^{\rm exp}_i(r,z,\phi)$ is the expected number of photo-electrons from the templates. The reconstructed vertex position is calculated by interpolating the $\chi^2$ distribution.

\par
To correct the vertex-dependent energy non-uniformity, spallation neutron data is used to model the corrections in terms of radius $r$ and $z$. The reconstructed energy is given by:
\begin{equation}
E^{SN}_{\rm rec} = C^{SN}_r(r)\times C^{SN}_z(z)\times Q_{\rm tot} / C^{SN}_e \,,
\end{equation}
where $C^{SN}_r(r)$ and $C^{SN}_z(z)$ are 3rd-order polynomials derived from the observed non-uniformity of the spallation neutron energy response in the AD, $Q_{\rm tot}$ is the observed total charge, and $C^{SN}_e$ is a charge to energy conversion coefficient determined by constraining the spallation neutron peak to 8.047 MeV.

\par
\begin{figure}[htb]
\centering
\includegraphics[width=7cm]{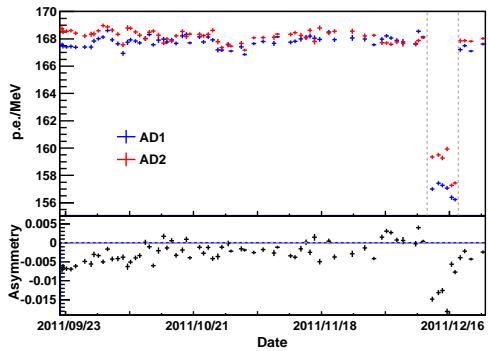}
\caption{Variation in the spallation neutron energy conversion coefficient with time is shown at top.  The associated Asymmetry parameter is shown at bottom.  The dashed vertical lines bound a period of dedicated systematic studies.}
\label{fig:spatvar}
\end{figure}
\begin{figure}[htb]
\centering
\includegraphics[width=7cm]{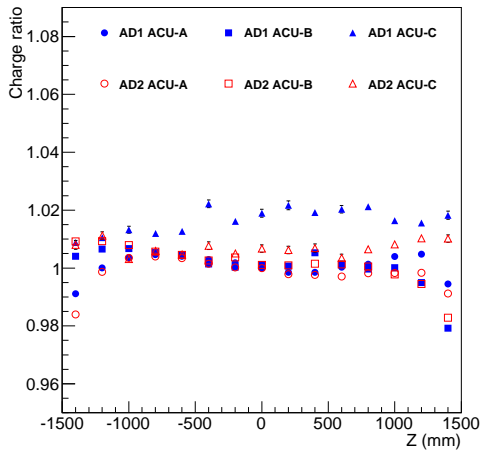}
\caption{Variation in the $^{60}$Co energy peak with source location after spallation neutron based vertex correction.}
\label{fig:spaco}
\end{figure}

The charge to energy conversion coefficient $C^{SN}_e$ is approximately 168 p.e./MeV for AD1 and 169 p.e./MeV for AD2. The variation of this calibration constant with time is shown in Fig.~\ref{fig:spatvar}. Although data from special runs are included in the plot ({\rm e.g.} PMT systematic studies) data from these special runs are not included in the side-by-side comparison.  Excluding these runs, the absolute variation is less than 1\%. The time dependence of the energy reconstruction is estimated by monitoring the $^{60}$Co calibration data. The reconstructed energy of the $^{60}$Co sources at the detector center varies 0.3\% for both ADs, while the rms of the Asymmetry drift over time is only 0.15\%.  The correction to the non-uniformity is also checked with the $^{60}$Co data, as shown in Fig.~\ref{fig:spaco}. After vertex correction, the energy non-uniformity is almost the same as in Fig.~\ref{fig:reczscan}.

\par
The energy reconstruction methods are consistent with each other, and each offers distinct advantages.  The source data method has the advantages of extremely low background, high statistics, and identicalness at near and far sites.  Furthermore, the source calibration data alone is sufficient to reconstruct the spatial uniformly distributed IBD neutron peak and the spallation neutron peak at their true energy, as we will see in later sections.  This implies that the detector response is well understood.

Spallation neutrons have a spatial distribution nearly as uniform as the IBD neutrons.  Therefore, this energy calibration method equally samples the entire target volume and corrects the spatial non-uniformity of IBD neutrons more accurately.  Furthermore, spallation neutron events are extracted from all regular physics runs, compared to the weekly source calibrations.

\par
\begin{figure}[htb]
\centering
\includegraphics[width=7cm]{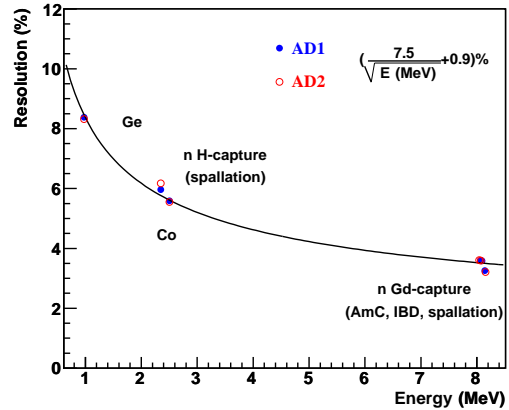}
\caption{Resolution of reconstructed energy.}
\label{fig:eres}
\end{figure}

The resolution of the reconstructed energy is shown in Fig.~\ref{fig:eres}.  The resolution curve is fitted phenomenologically to be (7.5/$\sqrt{E({\rm MeV})}+0.9)\%$.

\subsection{AD triggers\label{sec:adt}}
\par
\begin{figure}[htb]
\centering
\includegraphics[width=7cm]{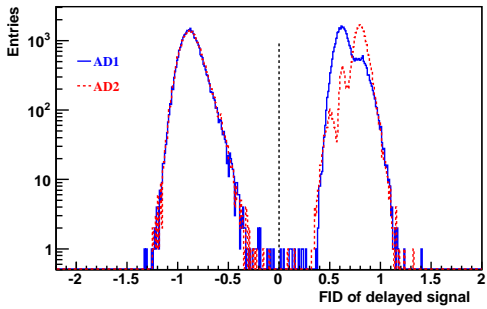}
\caption{Flasher event discrimination for the delayed signal of IBD candidates. Events with a PMT discharge (flasher) have a FID $>0$ while real IBD events have a FID $<0$.}
\label{fig:flasher}
\end{figure}

AD data is primarily collected through an OR of the multiplicity and energy sum triggers with an energy threshold around 0.4 MeV (see Fig. 9).  The typical trigger rate is about 280 Hz per AD.
About 5\% of these are so-called flasher events, instrumental background events apparently resulting from an electronic discharge in the PMT assembly.  The observed energy of flasher events ranges from threshold to 100 MeV.
About 5\% of the PMTs have been identified as sources of flasher events. The flasher events have a geometric charge pattern that is distinct from other physical events.  As such, they can be easily identified and removed.  Fig.~\ref{fig:flasher} shows the discrimination of flasher events for the delayed signal of IBD candidates, where the discrimination variable FID $>0$ indicates a flasher. For this demonstration, flasher events have been kept in the IBD selection.  For the IBD analysis as well as most other analyses, the rejection of flasher events will be done at the beginning of the data reduction. Most of the events around FID $=0$ are actually pile-up events that confuse the discrimination algorithm.  The distributions for AD1 and AD2 overlay well for the selected IBD events (FID $<$ 0), but there is some variation between the distributions when FID $>$ 0.  Monte Carlo studies indicate that the FID is extremely efficient at identifying flasher events. The inefficiency for selection and contamination of IBD selection are evaluated to be 0.02\% and $<10^{-4}$ respectively, for both AD1 and AD2. The relative uncertainties are estimated to be 0.01\%. Special runs were conducted to identify PMTs that exhibit flashing; however, due to the high efficiency of the FID, 100\% of the AD PMTs are operational, including the PMTs identified as flashers.

\par
\begin{figure}[htb]
\centering
\includegraphics[width=7cm]{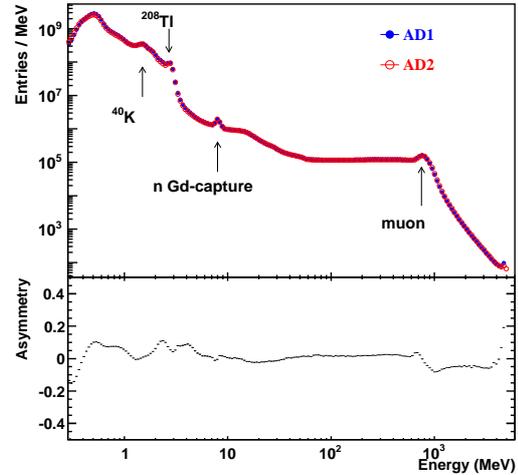}
\caption{Spectrum comparison of all AD triggers for AD1 and AD2 after flasher rejection. The lower plot measures the Asymmetry between AD1 and AD2, which is largely due to nuisance triggers immediately following a muon event.}
\label{fig:alltrig}
\end{figure}

The spectra of AD triggers for both detectors are shown in Fig.~\ref{fig:alltrig} after removing flasher events, along with the associated Asymmetry parameter.  The knee at $\sim$$10^3$ MeV corresponds to the maximum path-length by a minimum ionizing muon through the 4-m  diameter by 4-m high active detector volume. Higher energy events are showers induced by muons. The greatest difference between the spectra, about $\sim$15\%,  is created in large part by triggers that occur shortly after a muon passes through a detector. These nuisance triggers may last up to 10 $\mu$s following a high-energy muon, and result from PMT after-pulsing and ringing, and signal overshoot in the readout chain~\cite{Soren}.

\par
\begin{figure}[htb]
\centering
\includegraphics[width=7cm]{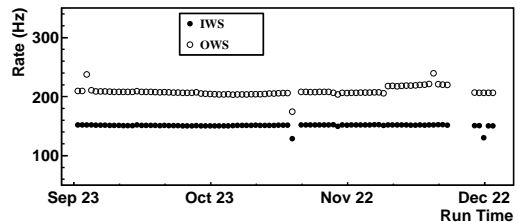}
\caption{Muon rate in the IWS and OWS with a PMT multiplicity $\geq$12.}
\label{fig:wsmuon}
\end{figure}

The performance of the Daya Bay muon system will be described elsewhere.
The event rates for a multiplicity of twelve or more PMTs in the IWS and OWS are shown in Fig.~\ref{fig:wsmuon}. There are 121 PMTs installed in the IWS and 167 PMTs in the OWS. Occasionally, the muon rates are disturbed by electronics noise. The observed rate of muons in each AD is 21 Hz, assuming that any AD trigger with an energy $>$20 MeV results from a muon.  For such an AD muon, the detection efficiency of the IWS is 99.7\%, and that of the OWS is $\sim$97\%.  We believe that these are lower limits since high energy AD events may be due to physical processes other than muons (e.g. from high energy neutrons produced in the surrounding rock). Such events have been observed in Monte Carlo simulations. The apparent inefficiency of the OWS is largely due to muon decay in the IWS or an AD.

\par
\begin{figure}[htb]
\centering
\includegraphics[width=7cm]{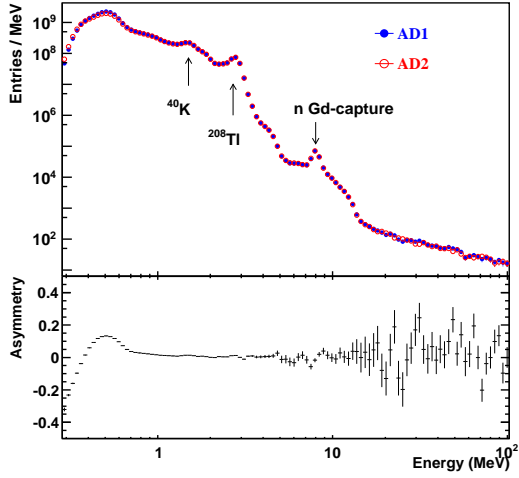}
\caption{Spectrum comparison of AD triggers after muon veto cuts have been applied.  The associated Asymmetry distribution is also shown.}
\label{fig:trigmuon}
\end{figure}

Figure~\ref{fig:trigmuon} displays the AD spectra after excluding AD triggers in the time interval (-2, 200) $\mu$s with respect to an IWS or OWS trigger. The -2 $\mu$s cut is used to avoid time alignment issues between detectors.  Besides the removal of AD events with muons or muon daughter products, nuisance triggers that closely follow AD muons are also rejected.  Good agreement is observed in the spectra between the ADs.

\subsection{IBD rates and energy spectra}

\par
To select IBD events, the following criteria are applied:
\begin{enumerate}
\item the flasher rejection cut described above;
\item positron energy: 0.7 MeV $<E_{\rm prompt}<$ 12.0 MeV;
\item neutron energy: 6.0 MeV $<E_{\rm delayed}<$ 12.0 MeV;
\item time coincidence: 1 $\mu$s $< \Delta t_{\rm prompt,delayed} <$ 200 $\mu$s;
\item a muon veto that rejects a prompt-delayed pair if the delayed signal is within 600 $\mu$s after an IWS or OWS trigger; however, if a muon deposits $>$~20~MeV in an AD the exclusion window is extended to 1000 $\mu$s due to the increased probability of multiple neutrons.  If a muon deposits $>$ 2.5 GeV in an AD, the exclusion window extends to 1.0 second to reject long-lived cosmogenic backgrounds.
\item a multiplicity cut that requires no other $>$ 0.7 MeV trigger 200 $\mu$s before the prompt signal and 200 $\mu$s after the delayed signal;
\end{enumerate}

\par
\begin{figure}[htb]
\centering
\includegraphics[width=7cm]{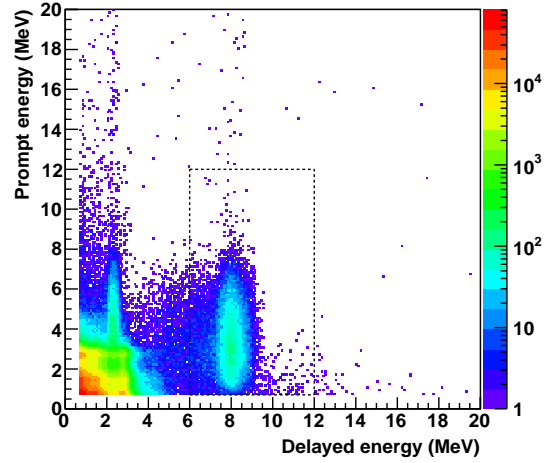}
\caption{Scatter plot of the prompt-delayed signals. Pairs inside the dotted line box are taken as IBD candidates.}
\label{fig:ibd2d}
\end{figure}

A scatter plot of prompt-delayed energy pairs is shown in Fig.~\ref{fig:ibd2d}.
In the plot, IBD candidates with neutron capture on hydrogen can also be identified although the low energy part is obscured by accidental coincidence background.

\par
Two detector-dependent inefficiencies, the muon veto dead time and the multiplicity cut, have been corrected in the following studies on the IBD rate and spectrum comparison. The energy selection, time coincidence selection, as well as other absolute IBD efficiencies such as the fraction of neutron capture on Gd are not corrected. They are identical for ADs from our analysis thus not relevant to the side-by-side comparison of ADs. However, the uncertainties are analyzed.

\par
All triggers within 2 $\mu$s before an IWS or OWS muon are also vetoed to avoid time alignment issues between detectors. IWS and OWS veto time intervals are carefully merged with the veto time interval of each AD to avoid double counting.   All livetime segments between muons are precisely calculated.  The total veto dead time is measured to be 17.68\% for AD1 and 18.06\% for AD2 where the difference comes from the AD response to high-charge events and resulting muon rejection..

\par
The multiplicity cut efficiency consists of three parts.
\begin{enumerate}
\item  No other $>$ 0.7 MeV trigger (singles) in a 200 $\mu$s window before the prompt signal. The efficiency is $\epsilon_1=1-R_s\cdot 200$ $\mu s$ where $R_s$ is the rate of singles, which is $\sim$60 Hz in both ADs.
\item No singles between the prompt and delayed signals.  The efficiency is $\epsilon_2=1-R_s\overline{T}_c$, where $\overline{T}_c$ is the average neutron capture time.
\item No singles in a 200 $\mu$s window after the delayed signal.  The efficiency is $\epsilon_3=1-R_s\cdot 200$ $\mu s(1-100$ $\mu s/T_v)$ for a single livetime segment $T_v$ longer than 200 $\mu$s, or $\epsilon_3=1-R_s T_v/2$ for $T_v$ shorter than 200 $\mu$s. $\epsilon_3$ is different from $\epsilon_1$ because a muon may occur and veto the following singles, thus reducing the rejection probability.
\end{enumerate}
The multiplicity cut efficiency is evaluated using Poisson statistics and the probability density function of the neutron capture time. The difference from the above approximation is $<$ 0.015\%. The inefficiency from the multiplicity cut is $\sim$2.5\%, depending on the singles rate. Combining the multiplicity cut and the muon veto, the total efficiency for these two cuts are 80.10\% for AD1 and 79.76\% for AD2.

\par
Another analysis has been done with a different multiplicity cut that requires:
\begin{enumerate}
\item Only one prompt candidate within 200 $\mu$s before a neutron candidate;
\item No other prompt candidate within 400 $\mu$s before the neutron candidate;
\item No other neutron candidate within 200 $\mu$s after the first neutron candidate;
\item At the same time, applying an additional muon veto 200 $\mu$s before each IWS or OWS muon.
\end{enumerate}
The last cut is required to avoid a high energy ($>$~6~MeV) prompt signal of an IBD being mistaken as a neutron candidate and forming an accidental background with a preceding background gamma while the real neutron signal is vetoed by a muon. This multiplicity cut has a fixed time window and thus is decoupled from the neutron capture time and the muon veto, with the cost of dropping $\sim$4\% more IBD events. The selection efficiency is $\exp(-R_s\cdot 400$ $\mu s) \exp(-R_d\cdot 200$~$\mu s)$ where $R_d$ is the rate of neutron candidate. The two multiplicity cuts have been compared and verify the efficiency calculations.

\par
For the purpose of this side-by-side comparison, we have evaluated the
three largest backgrounds in the IBD sample: accidentals, $^9$Li/$^8$He decays, and fast neutrons.  Other correlated backgrounds such as ($\alpha$,n) reactions or the ACU sources have been evaluated and found to be negligible.

\par
The accidental background rate was determined by separately counting the singles rate for both $e^+$-like and neutron-like signals. We get $10.20\pm0.05$/day for AD1 and $10.10\pm0.05$/day for AD2, corresponding to $\sim$1.7\% Background-Signal ratio (B/S) and a 0.01\% uncertainty in the analyzed data set.

\par
The correlated background from the $\beta$-$n$ cascade of $^9$Li/$^8$He decays is evaluated by fitting the time between the last muon and prompt IBD candidate~\cite{wenljnim} with the known decay times for these isotopes. The $^9$Li/$^8$He detection rate is found to be $4.2\pm1.2$/day.
The $^9$Li/$^8$He detection rates for muons that deposit $<$ 2 GeV in an AD is consistent with zero within statistical error.  After applying the 1-second shower muon veto, the $^9$Li/$^8$He background in the IBD sample is estimated to be $<$ 0.3\% (1 $\sigma$).

\par
The fast neutron backgrounds are estimated in two ways.
\begin{enumerate}
\item The upper limit of the prompt energy cut is relaxed to extend the prompt energy spectrum to high energy.  A flat distribution is observed for energies $>$ 12 MeV. Assuming the energy spectrum for fast neutrons is flat through the relevant neutrino energy region, we estimate the relative rate of the fast neutron background within the IBD sample to be 0.2\%.
\item For muons tagged by the IWS and OWS, we get the rate and energy spectrum of the fast neutron backgrounds as a function of muon track length in the water pool. A fast neutron can leak into the IBD sample if its parent muon is not detected due to increased inefficiency for detecting muons of short track length in the water pool.  Estimates of the fast neutron backgrounds produced by muons passing through nearby rock rely on simulations. The B/S is measured to be $0.15 \pm 0.05\%$, consistent with the first method.
\end{enumerate}

\par
For the side-by-side comparison of AD1 and AD2 in the same water pool, the background contents are very similar thus not critical for the relative comparison.

\par
\begin{figure}[htb]
\centering
\includegraphics[width=7cm]{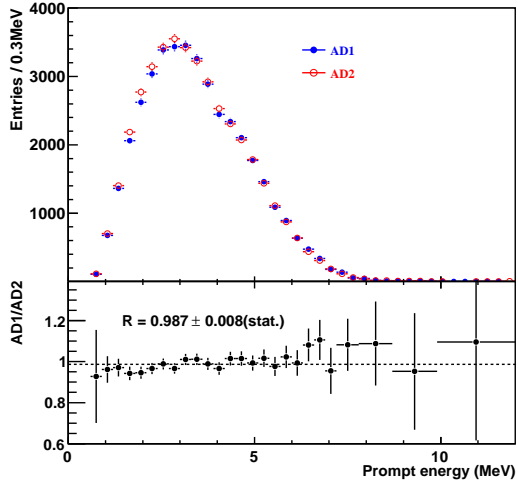}
\caption{Energy spectrum for the prompt signal of IBD events in each AD along with the ratio in each energy bin. The dashed line shows the ratio $R$ of the total rates in AD1 and AD2.}
\label{fig:nuspec}
\end{figure}

The prompt energy spectra of selected IBD events are shown in Fig.~\ref{fig:nuspec} for AD1 and AD2. The backgrounds, dominated by accidental coincidence, are subtracted statistically in the plot. The ratio of the total IBD rates in AD1 and AD2 is $0.987\pm0.008$ (stat).

\par
\begin{figure}[htb]
\centering
\includegraphics[width=7cm]{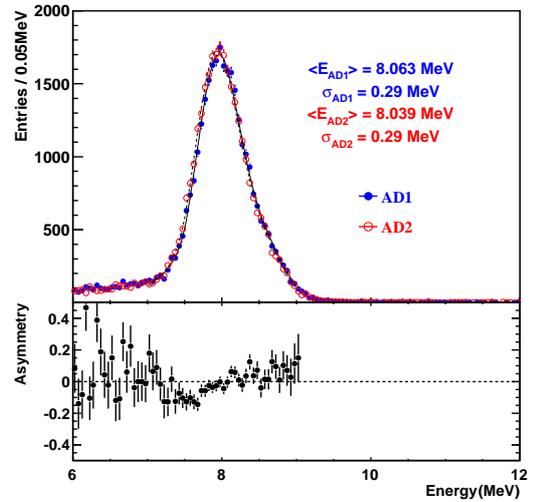}
\caption{IBD neutron energy distribution for each AD along with the associated Asymmetry distribution.}
\label{fig:ibdneutpeak}
\end{figure}

The neutron energy peak is a critical check on the energy scale calibration and related uncertainties. The neutron energy distribution of the selected IBD samples is shown in Fig.~\ref{fig:ibdneutpeak}. When fitted with double Crystal Ball functions described in Sec.\ref{sec:source}, the neutron peak agrees very well with the expected value of 8.047 MeV and the expected resolution. The peak for AD1 is $\sim$0.3\% higher than that for AD2, and will be discussed in Sec.~\ref{sec:eunc}.

\par
\begin{figure}[htb]
\centering
\includegraphics[width=7cm]{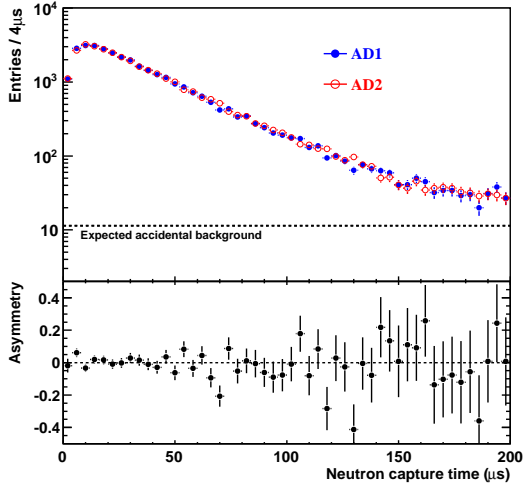}
\caption{Neutron capture time for the selected IBD sample along with the associated Asymmetry distribution.}
\label{fig:ibdtime}
\end{figure}

The capture time spectrum has four components. The neutrons from the IBD reactions thermalize in $\sim$8 $\mu$s  forming the rise in the spectrum.  Thereafter,  the capture on Gd dominates and forms the exponential component. At times larger than 100 $\mu$s, the spill-in/spill-out effects\footnote{'Spill-in'  ('spill-out') occurs when an IBD neutron produced outside (inside) the Gd-LS volume is detected in (outside) the Gd-LS volume. These phenomena effectively alter the target mass for antineutrinos.} at the edge of the target vessel play an important role.  Neutrons produced in the gamma-catcher survive for longer periods of time, allowing them an opportunity to spill-in to the target volume before being captured on Gd. Finally, there is a small flat component from accidental coincidences. The neutron capture time on Gd for the whole IBD sample is shown in Fig.~\ref{fig:ibdtime}. The expected accidental background component is also shown. To compare with the Am-$^{13}$C results, the sample is purified by applying a vertex cut ($r<$ 1.25 m and $|z|<$ 1.25 m) on the delayed signal events to reject spill-in candidates, and a cut on prompt energy ($E_{\rm prompt}>3$ MeV) to further reject the accidental backgrounds. The neutron capture times for the cleaned IBD samples are $28.2\pm0.3$ $\mu$s for AD1 and $28.6\pm0.3$ $\mu$s for AD2, in good agreement with the Am-$^{13}$C results.

\par
\begin{figure}[htb]
\centering
\includegraphics[width=\columnwidth]{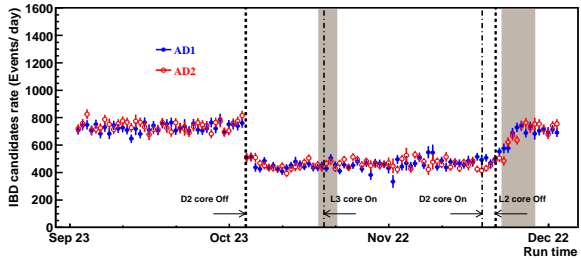}
\caption{IBD rate in AD1 and AD2 as a function of time.}
\label{fig:ibdrate}
\end{figure}

The IBD rates are shown in Fig.~\ref{fig:ibdrate} as a function of time. The D2 core was shut down on October~26, 2011 and returned to service on December~9.  Normally, reactor cores are ramped-up over 1-2 weeks before reaching full power. At full power, the D2 core contributes around 40\% of the neutrinos observed by AD1 and AD2.  The ramping can clearly be seen in the observed IBD rate. On November~9, the L3 core came back online after refueling.  On December~12, the L2 core was shutdown. The L2 and L3 cores contribute $\sim$8\% and $\sim$3\% of the neutrinos observed by AD1 and AD2, respectively.  During the three-months of data taking, there were two periods (shown as vertical shaded areas) amounting to 203.2 hours that were dedicated to test runs ({\em e.g.}~flasher and PMT testing).

\section{Detection uncertainties}
\par

\subsection{Target mass}
\par

The target mass is defined as the mass of Gd-LS inside the 3-m inner acrylic vessel (IAV), up to the top of its conical lid. Two sets of bellows extend above the IAV, connecting the IAV to the central overflow tank and a fluid monitoring camera.  The target mass is determined from the total mass of Gd-LS added to the detector during filling, minus the mass of the fluid in the overflow tanks and the two bellows. The fluid height  in the overflow tank is monitored by a multiply-redundant system of sensors.  The mass of Gd-LS in the bellows is computed from the bellows volume and the Gd-LS density.

The total number of free protons (hydrogen nuclei) is simply given by the target mass multiplied by the number of hydrogen nuclei per kg of Gd-LS.  The Gd-LS mass density is described by the formula
\begin{equation}
\label{eqn:density}
\rho_t=\rho_0/\left(1+\beta(T-T_0)\right),
\end{equation}
where $\rho_0=0.8590$~kg/L at $T_0=19$~C and $\beta=9.02\pm0.16 \times 10^{-4}$.
Based on a purely analytical calculation incorporating a realistic distribution of LAB hydrocarbon chains, the proton mass fraction in Gd-LS is determined to be 11.77\%.  A combustion-based measurement of the hydrogen fraction in Gd-LS  is used in this analysis and gives $12.01\pm0.42$\% and $11.97\pm0.47$\% in ADs 1\&2, respectively, including an instrumental uncertainty of 0.3\%.   Since the measurements and calculation are consistent, and the ADs are filled equally from five Gd-LS storage tanks, we consider the relative uncertainty on the free proton mass fraction between detectors negligible. For the absolute mass fraction uncertainty, we take the largest uncertainty, 0.47\%.

There are no ribs or other structures inside the 3m vessel to trap bubbles. The IAV is a vertical right cylinder with inner diameter 3.10 m and inner height 3.07 m. The cylinder is topped by a shallow cone with diameter 3.04 m and acute 3 degree edge angle. The total volume is 23.364 m$^3$.  Multiplying by a density of 860 kg/m$^3$ gives a nominal target mass of 20,093 kg. Since the blinding is 120 kg, a nominal target mass of 20,000 kg is assumed for simplicity.

The total mass of Gd-LS inside the AD is determined during filling. This quantity is subsequently blinded at the
level of 0.6\% (120 kg).  The mass of Gd-LS is determined primarily from load cells under the ISO tank, with a backup measurement from a Coriolis flow meter.  Multiple calibrations during ISO tank filling indicate that the four load cells are linear.   Long-term tests with the load cells show that the best precision is obtained from a 15-minute average of load cell readings, and that the load cell readings may drift by up to 3 kg over several hours.  Corrections are made to account for the weight of dry-nitrogen that enters the ISO tank during filling.  The mass as determined by the Coriolis meter is consistent within the accuracy of the meter for both ADs. Overall, the relative uncertainty on the total mass is dominated by the irreducible load cell drift.  The combined uncertainty on the total mass is 3.0 kg or 0.015\%.

The overflow tanks at the top of the AD accommodate fluid expansion and contraction from temperature changes. The mass in the overflow tanks is computed from the liquid levels, the tank geometry, and the liquid density. The liquid level is determined from a redundant set of lid sensors. The most accurate sensor uses ultrasonic sound to measure the distance from the sensor to the fluid with 1 mm accuracy. Cross-checks on this sensor are provided by a capacitance sensor, and by camera observation of the liquid levels in the off-center calibration port. Although the ultrasonic sensors have a resolution of 0.1 mm, calibration reveals deviations from linearity at the 1-mm level. A 1 mm uncertainty in liquid height corresponds to a 1.33 L, or 1.14 kg (0.0057\%) uncertainty on the target mass. The conversion from liquid height to liquid volume is calculated from a survey of the overflow tank geometry, and was cross-checked by filling one of the tanks with deionized water in 2L increments. The largest difference between calculated and measured liquid volume is 1.5 L. Multiplying by a fluid density of 0.86 kg/L gives an uncertainty of 1.3 kg. Surveys of all overflow tanks show good similarity, so we use a general function to compute the overflow mass of all tanks with a maximum deviation from any physical tank of 0.22 kg. Combining these two uncertainties gives a total uncertainty on the calculation of fluid mass from fluid height of 1.32 kg or 0.0066\%.  A tilted overflow tank would cause an error in fluid height as measured by the offcenter ultrasonic sensor.  The levelness of the ADs has been measured, and the uncertainty from the tilt is less than 1.37 kg, or 0.0068\%.

Two identical bellows connect the 3m inner acrylic vessel to the SSV lid. The bellows are corrugated, with cylindrical cuffs on each end. The bellows do not compress during AD assembly, but the cuffs are somewhat free to slide in their housing. Assuming relatively large uncertainties on the bellows dimensions, 5\% cross-section and 8\% (6 cm) length, the bellows volume is $4.30 \pm 0.42$ L or $3.70 \pm 0.36$ kg.  The bellows are attached to the IAV lid at two stubs. The total fluid volume in the two stubs is 5.78 L. The volume uncertainty is assumed to be equivalent to the bellows uncertainty, 0.4 L. Overall, the bellows and stub uncertainty is 0.58 L or 0.5 kg (0.0025\%).  The uncertainties on the number of free protons are summarized in Table~\ref{tbl:uncertainties}.

\begin{table}[htb]
\caption{Uncertainties on the number of target protons.  Relative and absolute uncertainties are given for a single AD.
Fractional uncertainties are computed assuming a nominal 20~t target mass.
\label{tbl:uncertainties}
}
\begin{tabular}[c]{l|l|l}
\hline
Quantity                        & Relative &  Absolute  \\
\hline
Free protons/kg                      & neg.     &  0.47\%    \\
Density (kg/L)                  & neg.     &  0.0002\%     \\
Total mass                      & 0.015\%  & 0.015\%   \\
Overflow tank geometry          & 0.0066\% &  0.0066\%  \\
Overflow sensor calibration     & 0.0057\% & 0.0057\%  \\
Overflow tank tilt              & 0.0068\% & 0.0068\%  \\
Bellows Capacity                & 0.0025\% &  0.0025\%  \\
\hline
Target mass                     & 0.019\%  & 0.019\%   \\
Free protons                  & 0.019\%  & 0.470\%   \\
\hline
\end{tabular}
\end{table}

\subsection{Energy scale uncertainty\label{sec:eunc}}

\par
As shown in Sec.~5.1, the energy scale uncertainty from the time variation is estimated to be $\pm$0.2\% by averaging the time variation of all ACUs. The non-uniformity differences between two ADs, when averaging ACU-A and ACU-B, lead to an uncertainty of $\pm$0.2\%. The non-linearity from the cobalt energy scale to the neutron energy scale should be the same for the two ADs.  We find that the difference is 0.1\%. From the comparison of calibration data, we estimate that the relative energy scale uncertainty for neutrons uniformly distributed in the target volume is $\sim$0.3\%.

\par
\begin{figure}[htb]
\centering
\includegraphics[width=7cm]{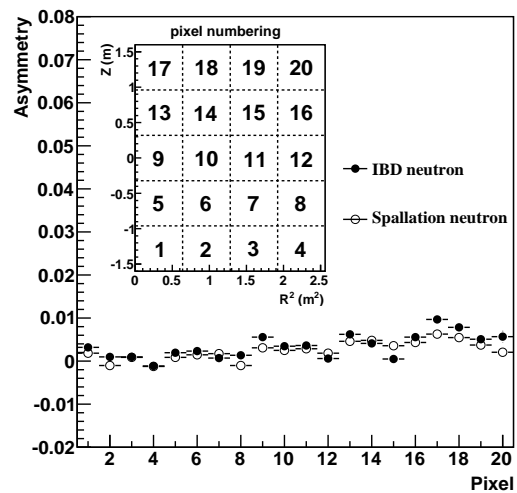}
\caption{Asymmetry distribution in the peak of the energy distribution shown bin by bin for spallation and IBD neutrons.}
\label{fig:neutmapcom}
\end{figure}
\begin{figure}[htb]
\centering
\includegraphics[width=7cm]{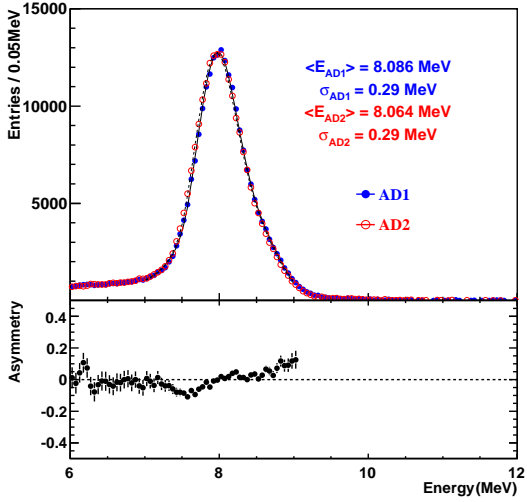}
\caption{Spallation neutron energy distribution for each AD along with the associated Asymmetry.}
\label{fig:spaneutpeak}
\end{figure}

Besides the calibration data, the energy scale uncertainty related to the non-uniformity can be evaluated from the IBD neutron and spallation neutron distributions. Fig.~\ref{fig:neutmapcom} displays the Asymmetry of energy peaks between the two ADs for both IBD and spallation neutrons captured on Gd. The target volume is divided into twenty bins (pixels), four along $R^2$ and five along the z direction. The indexing of the bins is shown in the figure. Each bin has equal volume thus equal numbers of neutrons, assuming the spatial distribution is uniform. The actual relative bin content could be slightly different due to the spill-in/spill-out effects, slowing down of spallation neutrons, vertex reconstruction and event selections.  Events reconstructed outside the target volume have been included into the closest bin. The Asymmetry for IBD neutrons ranges from -0.1\% to 1.0\%, with a fitting error of 0.1\%. Spallation neutrons have the same structure as IBD neutrons, verifying the non-uniformity differences between two ADs. The mean values of the Asymmetry in the twenty bins are 0.3\% for both IBD neutrons and spallation neutrons. By the definition of the Asymmetry, it means that the source-based calibration results in a  0.3\% higher energy scale for AD1 than AD2, agreeing with the aggregate energy peak difference shown in Fig.~\ref{fig:ibdneutpeak} and Fig.~\ref{fig:spaneutpeak}. The difference appears in the upper half of the ADs. For the same reason, the energy Asymmetry of any uniformly distributed sources, such as alphas in Fig.~\ref{fig:nonlin}, will be 0.3\% higher than the Asymmetry of sources at the detector center. The RMS value of the bins is 0.27\% for IBD neutrons and 0.21\% for spallation neutrons, which will be taken as the relative energy scale uncertainties from the non-uniformity.

\par
The direct comparison of the IBD neutron energy has been shown in Fig.~\ref{fig:ibdneutpeak}. The difference between the two ADs is 0.3\% with AD1 being higher. For spallation neutrons, the same comparison is shown in Fig.~\ref{fig:spaneutpeak}. The difference between the two ADs is also 0.3\% with AD1 being higher, but systematically higher than IBD neutrons also by 0.3\%.  Combining the time variation, the non-uniformity, and the non-linearity uncertainties, we estimate that the relative energy scale uncertainty for the 8.047 MeV neutron peak for this pair of ADs is 0.4\%.  This calibration-related uncertainty is reflected in the 0.3\% relative energy scale difference observed after source-based calibration.

\par
The efficiency of the 6 MeV cut for the neutron selection (Sec.5.5) suffers from edge effects in the ADs. The neutrons at the detector center are not affected by the 6 MeV cut selection since the energy is fully contained, as shown in Fig.~\ref{fig:neutronfit}. When approaching the detector edge, the non-uniformity Asymmetry between the ADs increases.  We use a high statistics IBD simulation to generate a weight map corresponding to the non-uniformity map in Fig.~\ref{fig:neutmapcom}, by counting the neutrons around 6 MeV. After re-weighting, the relative energy scale uncertainty for the 6 MeV cut is estimated to be 0.45\% for this pair of ADs. The relative efficiency uncertainty for the 6 MeV neutron energy cut, when the energy scale varies by 0.45\%, is 0.11\%.

\subsection{Summary of detector-related uncertainties}

\par
Relative uncertainties from other analysis cuts applied in the IBD selections are summarized below.
\begin{enumerate}
 \item The flasher rejection cut is rather safe and robust for the IBD selections as shown in Sec.~\ref{sec:adt}. The uncertainty is estimated to be 0.01\%.
 \item The AD trigger has almost 100\% efficiency at 0.7 MeV, which is 3$\sigma$ away from the $e^+$ IBD threshold. The IBD reaction in acrylic may have prompt energy lower than 0.7 MeV. Assuming a 2\% energy scale uncertainty to account for the edge effect, we get 0.01\% uncertainty for the $E_{\rm prompt}$ selection efficiency by Monte Carlo simulation.
 \item In principle, the neutron capture time for the two ADs should be identical due to the specially designed Gd-LS mixing and filling procedure. From the fitted Am-$^{13}$C neutron capture time, IBD neutron, and spallation neutron capture time, the Gd concentration difference is estimated to be $<$ 0.5\%. By varying the Gd concentration in Monte Carlo, we find the uncertainty of the time coincidence cut to be $<$ 0.01\%.
 \item Of the three components of the multiplicity cut, $\epsilon_1$ and $\epsilon_3$ can be accurately calculated, up to the statistical precision of singles. $\epsilon_2$ relies on the average neutron capture time. If the Gd content in the ADs is the same, the efficiency uncertainty is negligible. By measuring the average neutron capture time, this uncertainty is estimated to be $<$ 0.01\%.
 \item  From neutron capture time of Am-$^{13}$C sources and IBD neutrons, we limit the relative fraction of neutron captures on Gd in the ADs to $<$ 0.1\%.
 \item Relative differences in acrylic vessel thickness and density, and liquid density between ADs can result in relative differences in the number of n-Gd captures originating from outside the target region.  Using MC simulation, the measured sub-percent relative differences in these quantities in AD1 and AD2 lead to an expected relative spill-in/out uncertainty of 0.02\%.
 \item The determination of the livetime includes precision calculations of the muon veto time
and blocked triggers. The precision of the livetime is estimated to be $<$~0.01\%.
\end{enumerate}

\par
The detector-related relative uncertainties for AD1 and AD2 are summarized in Table~\ref{tab:keynumber}. The efficiencies of the multiplicity cut and capture time cuts, the relative precision on H/Gd ratio, and the relative uncertainties of spill-in/out effects are correlated; however, the H/Gd ratio uncertainty is dominant among the four. The total uncertainty is estimated to be 0.2\%.

\begin{table}[htb]
\begin{center}
\begin{tabular}[c]{ll} \hline
 Source of uncertainty & Quantity \\ \hline\hline
 Mass measurement relative precision  & 0.02\% \\
 Flasher cut & 0.01\% \\
 Efficiency of neutron energy cut & 0.11\% \\
 Efficiency of $e^+$ threshold cut & 0.01\% \\
 Efficiency of multiplicity cut & $<$ 0.01\% \\
 Efficiency of capture time cuts & 0.01\% \\
 Relative precision on H/Gd ratio & $<$0.1\% \\
 Relative uncertainty of spill-in/out & 0.02\% \\
 Livetime precision & $<$ 0.01\% \\ \hline\hline
 Total detector-related uncertainty & 0.2\% \\\hline
\end{tabular}
\caption{Detector-related relative uncertainty of Daya Bay evaluated with the 1st pair of ADs in the Daya Bay Near Hall.} \label{tab:keynumber}
\end{center}
\end{table}

\par
When the analyses were frozen, the baselines, the thermal power histories of the cores, and the target masses of the two ADs were unblinded.  The expected ratio of the IBD rates in AD1 and AD2 is 0.981 as compared to the measured ratio $0.987\pm0.008 {\rm (stat)}\pm 0.003 {\rm (syst)}$. The deviation from unity of this ratio is largely due to small differences in the baselines of the two ADs. The reactor flux differences can be ignored, and target mass difference contributes 0.15\% to the deviation.

\section{Conclusion}
\par
Controlling systematic uncertainties in order to measure sin$^22\theta_{13}$ with a sensitivity better than 0.01 at the 90\% confidence level is a challenge. A relative measurement using near and far antineutrino detectors can greatly reduce the detector-related systematic uncertainties. The Daya Bay experiment is designed with eight nearly identical antineutrino detectors. Multiple antineutrino detectors at each site enable side-by-side comparisons to estimate the relative uncertainties in detector efficiencies. The first two antineutrino detectors have been installed in EH1 and are operating with steady data-taking since September 23, 2011. The analysis of the first three months' data demonstrates a comprehensive understanding of the detector responses and the relative detection efficiencies. The relative energy scale uncertainty is determined to be 0.4\% and the relative efficiency for the neutron energy selection is 0.11\%. Combining with the precise target mass measurement and selection efficiencies, the relative neutrino detection efficiency uncertainty is 0.2\% for the first two antineutrino detectors, an improvement over the design value of 0.38\%~\cite{dyb}. In this analysis,
the expected ratio of the IBD rates in AD1 and AD2 is 0.981, compared to the measured ratio $0.987\pm0.008 {\rm (stat)}\pm 0.003 {\rm (syst)}$.

\section{Acknowledgment}
\par
The Daya Bay experiment is supported in part by the Ministry of Science and Technology of China (contract no. 2006CB808100), the United States Department of Energy (DE-AC02-98CH10886, DE-AS02-98CH1-886, DE-FG05-92ER40709, DE-FG02-07ER41518, DE-FG02-84ER-40153, DE-FG02-91ER40671, DE-FG02-08ER41575, DE-FG02-88ER40397, DE-SC0003915, and DE-FG02-95ER40896),  the Chinese Academy of Sciences (KJCX2-EW-N08), the National Natural Science Foundation of China (Project number 10890090), the Guangdong provincial government (Project number 2011A090100015), the Shenzhen Municipal government, the China Guangdong Nuclear Power Group, Shanghai Laboratory for particle physics and cosmology, the Research Grants Council of the Hong Kong Special Administrative Region of China (Project numbers 2011A090100015, 400805, 703307, 704007 and 2300017), the focused investment scheme of CUHK and University Development Fund of The University of Hong Kong, the MOE program for Research of Excellence at National Taiwan University and NSC funding support from Taiwan, the U.S. National Science Foundation (Grants PHY-0653013, PHY-0650979, PHY-0555674, PHY-0855538 and NSF03-54951), the Alfred P. Sloan Foundation, the  University of Wisconsin, the Virginia Polytechnic Institute and State University, Princeton University, California Institute of Technology, University of California at Berkeley,  the Ministry of Education, Youth and Sports of the Czech Republic (Project numbers MSM0021620859 and ME08076), the Czech Science Foundation (Project number GACR202/08/0760), and the Joint Institute of Nuclear Research in Dubna, Russia.  We are grateful for the ongoing cooperation from the China Guangdong Nuclear Power Group and China Light \& Power Company.

\end{document}

%% file: dyb_authors_AD12.tex
%\author[]{Daya~Bay~Collaboration}

\author[IHEP]{F.~P.~An}
\author[USTC]{Q.~An}
\author[IHEP]{J.~Z.~Bai}
\author[UW]{A.~B.~Balantekin}
\author[UW]{H.~R.~Band}
\author[BNL]{W.~Beriguete}
\author[BNL]{M.~Bishai}
\author[NUU]{S.~Blyth}
\author[BNL]{R.~L.~Brown}
\author[IHEP]{G.~F.~Cao}
% ...

\address[IHEP]{Institute~of~High~Energy~Physics, Beijing}
\address[USTC]{University~of~Science~and~Technology~of~China, Hefei}
\address[UW]{Department of Physics, University~of~Wisconsin, Madison, WI}
\address[BNL]{Brookhaven~National~Laboratory, Upton, NY}
\address[NUU]{National~United~University, Miao-Li}
% ...

%\fntext[deceased]{Deceased}